\documentclass[a4paper,fleqn,usenatbib]{mnras}

\usepackage{newtxtext,newtxmath}

\usepackage[T1]{fontenc}
\usepackage{ae,aecompl}

\usepackage[british]{babel}

\usepackage{graphicx}	
\usepackage{amsmath}	

\usepackage{siunitx}
\DeclareSIUnit\solarMass{\mbox{$M_\odot$}}

\title{Modelling neutron star mountains}

\author[F. Gittins, N. Andersson and D.~I. Jones]{
F. Gittins,\thanks{E-mail: f.w.r.gittins@soton.ac.uk}
N. Andersson
and D.~I. Jones
\\
Mathematical Sciences and STAG Research Centre, 
University of Southampton, Southampton SO17 1BJ, United Kingdom
}

\date{Accepted 2020 November 18. Received 2020 November 17; in original form 
2020 October 6}

\pubyear{2020}

\begin{document}
\label{firstpage}
\pagerange{\pageref{firstpage}--\pageref{lastpage}}
\maketitle

\begin{abstract}
As the era of gravitational-wave astronomy has well and truly begun, 
gravitational radiation from rotating neutron stars remains elusive. Rapidly 
spinning neutron stars are the main targets for continuous-wave searches 
since, according to general relativity, provided they are asymmetrically 
deformed, they will emit gravitational waves. It is believed that detecting such 
radiation will unlock the answer to why no pulsars have been observed to spin 
close to the break-up frequency. We review existing studies on the maximum 
mountain that a neutron star crust can support, critique the key assumptions and 
identify issues relating to boundary conditions that need to be resolved. In 
light of this discussion, we present a new scheme for modelling neutron star 
mountains. The crucial ingredient for this scheme is a description of the 
fiducial force which takes the star away from sphericity. We consider three 
examples: a source potential which is a solution to Laplace's equation, another 
solution which does not act in the core of the star and a thermal pressure 
perturbation. For all the cases, we find that the largest quadrupoles are 
between a factor of a few to two orders of magnitude below previous estimates of 
the maximum mountain size.
\end{abstract}



\section{Introduction}

Neutron stars have long been of interest in gravitational-wave astronomy 
\citep{1978MNRAS.184..501P, 1984ApJ...278..345W}. This is owed to their extreme 
compactness (rivalled only by black holes) and their role in some of the most 
cataclysmic events in the Universe. There are a variety of mechanisms through 
which neutron stars can radiate gravitational waves. These include binary 
inspiral and merger \citep{2010CQGra..27q3001A}, various modes of oscillation 
\citep*[and their corresponding instabilities;][]
{1998ApJ...502..708A, 1999ApJ...516..307A} and rotating neutron stars 
deformed away from axial symmetry \citep{1998ApJ...501L..89B}. Recently, binary 
neutron stars have been the subject of significant interest since their 
exciting, first detections with gravitational-wave interferometers 
\citep{2017PhRvL.119p1101A, 2020ApJ...892L...3A}. This has reinvigorated the 
effort to explore other neutron star gravitational-wave scenarios. 

An open problem in the study of spinning neutron stars relates to the fact that 
no neutron star has been observed that spins (even remotely) close to the 
centrifugal break-up frequency -- which is generally above 
$\sim \SI{1}{\kilo\hertz}$ for most equation-of-state candidates 
\citep{2007PhR...442..109L}. The fastest discovered spinning pulsar rotates at 
$\SI{716}{\hertz}$ \citep{2006Sci...311.1901H}, well below this mass-shedding 
limit, and the current theory predicts that these stars reach these high 
frequencies through accretion, which should have no difficulty in spinning the 
neutron stars up to this limit \citep*{1994ApJ...424..823C}. It has been 
suggested 
that the lack of neutron stars spinning at these high rates is due to the 
emission of gravitational radiation which provides a braking torque that halts 
spin-up \citep{1998ApJ...501L..89B, 1999ApJ...516..307A, 2019MNRAS.488...99G}. 
The associated (quadrupole) deformations are commonly referred to as 
\textit{mountains}. 

Rapidly rotating neutron stars have, in fact, enjoyed the attention of a large 
number of searches using gravitational-wave data. These searches have been split 
into two strategies: looking for evidence of gravitational radiation for 
specific pulsars 
\citep{2004PhRvD..69h2004A, 2005PhRvL..94r1103A, 2007PhRvD..76d2001A, 
2008ApJ...683L..45A, 2010ApJ...713..671A, 2011PhRvD..83d2001A, 
2011ApJ...737...93A, 2014ApJ...785..119A, 2015PhRvD..91b2004A, 
2015PhRvD..91f2008A, 2017PhRvD..95l2003A, 2017PhRvD..96l2006A, 
2017ApJ...839...12A, 2017ApJ...847...47A, 2018PhRvL.120c1104A, 
2019PhRvD..99l2002A, 2019ApJ...879...10A, PhysRevD.100.122002} 
and wide-parameter surveys for unobserved sources 
\citep{2005PhRvD..72j2004A, 2007PhRvD..76h2001A, 2008PhRvD..77b2001A, 
2009PhRvD..79b2001A, 2012PhRvD..85b2001A, 2013PhRvD..87d2001A, 
2016PhRvD..94d2002A, 2017PhRvD..96f2002A, 2018PhRvD..97j2003A}. 
There has also been a study looking for gravitational waves from supernova 
remnants \citep{2019ApJ...875..122A}. The recent wide-parameter search 
\citep{2019PhRvD.100b4004A} has excluded the presence of fast-spinning neutron 
stars within $100$ parsecs with ellipticities larger than $10^{-8}$ and, most 
recently, the ellipticities of a number of observed pulsars have been 
constrained to less than $10^{-8}$ \citep{2020arXiv200714251T}. For these 
reasons, it is of great interest to calculate the largest mountain that a 
neutron star crust can sustain. This would provide an upper limit on the 
magnitude of gravitational-wave emission from these systems. 

There have been a number of studies of the maximum quadrupole 
deformation of a neutron star. The earliest of these was conducted by 
\citet*{2000MNRAS.319..902U}, who used the Cowling approximation in Newtonian 
gravity to derive an integral expression for the quadrupole moment. They 
introduced the argument that the body will obtain its maximum mountain when the 
crust is strained to its elastic yield point. This argument enabled them to 
straightforwardly find the strain tensor that ensures that every point in the 
crust is maximally strained. \citet*{2006MNRAS.373.1423H} observed that the 
approach of \citet{2000MNRAS.319..902U} did not respect the required boundary 
conditions at the base and top of the crust and that the Cowling approximation 
could have a large impact on the results. Therefore, they presented a 
perturbation formalism that relaxed the Cowling approximation and enabled them 
to treat the phase transitions appropriately. However, there are inconsistencies 
in their analysis which we explain later. The most recent estimates of the 
maximum elastic deformation have been provided by \citet{2013PhRvD..88d4004J}, 
who carried out their calculation in full relativity using a Green's function 
method. However, since they used the covariant analogue to the strain tensor 
from \citet{2000MNRAS.319..902U}, their calculation also ignored the boundary 
conditions on the crust. An important aspect of past studies is the 
fact the maximum mountains they calculate are independent of the precise 
mechanisms which sourced them. They do not consider the deforming forces or 
evolutionary scenarios which lead to the formation of the mountains.

We return to this problem to address some of the assumptions of the previous 
work and detail a formalism which enables one to accurately compute the 
quadrupole deformation throughout the star. As we show, in order to satisfy the 
necessary boundary conditions of the problem, it is extremely helpful to 
characterise the source of the perturbations. This has not been done in past 
calculations. In addition, it is not clear whether strain configurations where 
the majority of the crust is maximally strained can actually be reached in a 
real neutron star. The largest realistic mountain may be significantly smaller. 
These points suggest that future progress on this subject will rely 
on evolutionary 
calculations that consider the complete formation of the mountains 
\citep{1998ApJ...501L..89B, 2000MNRAS.319..902U, 2020MNRAS.493.3866S, 
2020MNRAS.494.2839O}. 

We begin, in Section~\ref{sec:Context}, with an introduction to static 
perturbations of neutron stars and a review of prior efforts on estimating the 
maximum mountain. We summarise their approaches and the important assumptions, 
which provide the motivation for this work. In Section~\ref{sec:Mountains}, we 
consider the necessary components of a neutron star mountain calculation. We 
provide a detailed discussion on the usual method of calculating mountains 
and introduce our own scheme, demonstrating the validity and equivalence of both 
approaches. We detail the perturbation formalism for our mountain scheme in 
Section~\ref{sec:Perturbation} and pay particular attention to the boundary 
conditions of the problem. We consider three sources for the deformations in 
Section~\ref{sec:Sources} and provide the maximum quadrupoles for each scenario. 
Finally, we conclude and discuss future work in Section~\ref{sec:Conclusions}. 

We adopt the usual Einstein summation convention where repeated indices indicate 
a summation. We use Latin characters $i, j, ...$ to denote spatial indices and 
use primes for derivatives with respect to the radial coordinate. We use 
$\delta$ and $\Delta$ to represent Eulerian and Lagrangian perturbations, 
respectively. These perturbations are related by 
$\Delta = \delta + \mathcal{L}_\xi$, where $\mathcal{L}_\xi$ is the Lie 
derivative along the Lagrangian displacement vector, $\xi^i$ 
\citep{1978ApJ...221..937F}. 

\section{Context}
\label{sec:Context}

When a star is deformed away from perfect sphericity it develops multipole 
moments. These are defined as 
\begin{equation}
    Q_{l m} \equiv \int_0^R \delta \rho_{l m}(r) r^{l + 2} dr, 
\label{eq:Multipole}
\end{equation}
where $(l, m)$ denotes the harmonic mode of the density perturbation, 
$\delta \rho_{l m}(r)$, and $R$ is the stellar radius. Note, in order to 
describe the full perturbative behaviour one would need to evaluate the sum 
over all modes, $\delta \rho(r, \theta, \phi) = \sum_{l = 0}^{\infty} 
\sum_{m = - l}^{l} \delta \rho_{l m}(r) Y_{l m}(\theta, \phi)$, where 
$Y_{l m}(\theta, \phi)$ are the usual spherical harmonics. However, since we are 
considering the quadrupole moment, $Q_{2 2}$, which is the dominant multipole in 
gravitational-wave emission, it is sufficient for the analysis to focus on the 
$(l, m) = (2, 2)$ mode. For this reason, we will drop the mode subscript on our 
perturbation variables.%
\footnote{One should note that, although we restrict ourselves to the 
$(l, m) = (2, 2)$ mode, other modes will contribute to the total strain, pushing 
the crustal lattice closer to the breaking strain, while not adding to the 
quadrupole.}

In addition to the quadrupole moment, we will quote our results using the 
fiducial ellipticity \citep{2005PhRvL..95u1101O}, 
\begin{equation}
    \epsilon = \sqrt{\frac{8 \uppi}{15}} \frac{Q_{2 2}}{I_{z z}}, 
\end{equation}
where $I_{z z}$ is the principal stellar moment of inertia, which we take to 
have the fiducial value of 
$I_{z z} = \SI{e45}{\gram\centi\metre\squared}$.%
\footnote{The fiducial principal moment of inertia can be different to the 
star's actual principal moment of inertia by a factor of a few.}
We do this to facilitate comparisons with observational papers. 

In this paper, we restrict ourselves to Newtonian gravity. Because of this, it 
is inappropriate to consider realistic equations of state and we assume a simple 
polytropic equation of state (Section~\ref{sec:Sources}). 

We consider perturbations of a non-rotating, fluid star with mass density 
$\rho$, isotropic pressure $p$ and gravitational potential $\Phi$. A barotropic 
fluid configuration, with velocity $v^i$, is a solution $(\rho, p, v^i)$ to the 
following equations: 
\begin{gather}
    \partial_t \rho + \nabla_i (\rho v^i) = 0, \label{eq:Continuity}\\
    \rho (\partial_t + v^j \nabla_j) v_i = - \nabla_i p 
        - \rho \nabla_i \Phi, \label{eq:Euler}\\
    p = p(\rho) \label{eq:EOS}
\end{gather}
and the gravitational potential is provided by Poisson's equation, 
\begin{equation}
    \nabla^2 \Phi = 4 \uppi G \rho. 
\label{eq:Poissons}
\end{equation}
Since the star is in equilibrium, the time derivatives vanish and, because it is 
static, we set $v^i = 0$ -- which means the continuity equation 
(\ref{eq:Continuity}) is trivially satisfied. 

To capture how the fluid elements move due to an induced perturbation, we 
introduce the Lagrangian displacement vector, $\xi^i$. This is related to the 
Lagrangian perturbation of the velocity \citep{1978ApJ...221..937F}, 
\begin{equation}
    \Delta v^i = \partial_t \xi^i. 
\end{equation}
The equations which govern the perturbations in a fluid are obtained by 
considering variations of equations~(\ref{eq:Continuity})--(\ref{eq:Poissons}). 
For a 
static background, $\delta v^i = \Delta v^i = \partial_t \xi^i$ and we have 
\begin{gather}
    \delta \rho + \nabla_i (\rho \xi^i) = 0, \label{eq:PerturbedContinuity}\\
    \rho \partial_t^2 \xi_i = - \nabla_i \delta p - \delta \rho \nabla_i \Phi 
        - \rho \nabla_i \delta \Phi, \label{eq:PerturbedEuler}\\
    \delta p = c_\text{s}^2 \delta \rho \label{eq:PerturbedEOS}
\end{gather}
and 
\begin{equation}
    \nabla^2 \delta \Phi = 4 \uppi G \delta \rho, 
\label{eq:PerturbedPoissons}
\end{equation}
where $c_\text{s}^2 \equiv dp/d\rho$ is the squared sound speed. Since we focus 
on static perturbations, we amend the perturbed Euler equation 
(\ref{eq:PerturbedEuler}) by 
\begin{equation}
    0 = - \nabla_i \delta p - \delta \rho \nabla_i \Phi 
        - \rho \nabla_i \delta \Phi + f_i, 
\label{eq:PerturbedEulerStatic}
\end{equation}
where $f_i$ is the density of a force which sustains the perturbations. The 
inclusion of this force enables us to produce non-spherical models and will 
prove to be an important component of our analysis, since it enables one to 
satisfy all the boundary conditions of the problem. We note that $f_i$ does not 
correspond to a physical force acting on the star. This force is a proxy for the 
(possibly quite complicated) formation history that results in its 
non-spherical shape. To study neutron stars with an elastic crust, we must 
modify (\ref{eq:PerturbedEulerStatic}) to include the shear stresses, 
\begin{equation}
    0 = - \nabla_i \delta p - \delta \rho \nabla_i \Phi 
        - \rho \nabla_i \delta \Phi + \nabla^j t_{i j} + f_i, 
\label{eq:PerturbedEulerElastic}
\end{equation}
where $t_{i j}$ is the shear-stress tensor, assumed to enter at the 
perturbative level. Here, we have used the same sign for the shear-stress tensor 
as in \citet{2000MNRAS.319..902U}. 

As we discuss in detail in Section~\ref{sec:Interface}, in order to connect the 
elastic crust of the star with the fluid regions, one needs to consider the 
traction vector. This must be continuous throughout the star. 

We turn our attention to past work on estimating the maximum mountain, which 
we now summarise. We do this to critique some of the assumptions made and set 
the stage for our new calculation. A convenient simplification that this body of 
work makes is to not (explicitly) consider the perturbing force. This is the 
main conceptual difference in our approach. We show how the force enters the 
problem in Section~\ref{sec:Mountains} and demonstrate that the formulation is 
consistent. 

\subsection{Ushomirsky, Cutler \& Bildsten}

The first (and perhaps most well known) maximum-mountain calculation was 
performed by \citet{2000MNRAS.319..902U}. They tackled the problem in Newtonian 
gravity and adopted the Cowling approximation -- neglecting perturbations of the 
star's gravitational potential, $\delta \Phi = 0$. The Cowling approximation 
means one can ignore perturbations in the fluid regions of the star, since the 
absence of shear stresses means there is no support for the pressure 
perturbations by the fluid [see (\ref{eq:PerturbedEulerStatic}) with $f_i = 0$]. 
Therefore, only perturbations in the crust contribute to the quadrupole moment. 

As is the current standard approach, \citet{2000MNRAS.319..902U} assumed that 
the crust manifests itself only at the perturbative level and, therefore, does 
not affect the equilibrium structure. From the perturbed Euler equation for 
an elastic solid [(\ref{eq:PerturbedEulerElastic}) with $f_i = 0$], they 
obtained an integral expression for the quadrupole moment of the star which 
depends on the shear stresses in the crust. In order to find an expression for 
these stresses, they conjectured that a star will attain its maximum quadrupole 
deformation when the crust is strained everywhere to the breaking point. To 
define the elastic yield limit, \citet{2000MNRAS.319..902U} used the von Mises 
criterion and further assumed that all the strain is in the $(l, m) = (2, 2)$ 
multipole. Thus, they analytically obtained the strain tensor which corresponds 
to the star being maximally strained. 

For a star with mass $M = \SI{1.4}{\solarMass}$ and radius 
$R = \SI{10}{\kilo\metre}$, \citet{2000MNRAS.319..902U} reported a maximum 
quadrupole moment of 
\begin{equation}
    Q_{2 2}^\text{max} \approx \num{1.2e39} \ 
        \left( \frac{\bar{\sigma}_\text{max}}{10^{-1}} \right) \ 
        \si{\gram\centi\metre\squared}, 
\label{eq:Q_22UCB}
\end{equation}
where $\bar{\sigma}_\text{max}$ is the breaking strain of the crust, which we 
take to have the canonical value, $\bar{\sigma}_\text{max} = 10^{-1}$ 
\citep{2009PhRvL.102s1102H}. In terms of the fiducial ellipticity, this result 
corresponds to $\epsilon^\text{max} \approx \num{1.6e-6} \ 
(\bar{\sigma}_\text{max} / 10^{-1})$. 

This approach, while elegant, does not enforce the continuity of the traction 
vector at the boundaries of the crust. At the base of the crust, there is a  
transition between the fluid core and the elastic crust. At the top, there is a 
transition between the elastic region and the fluid ocean. At these interfaces, 
there is expected to be a first-order phase transition where the crust sharply 
obtains a non-zero shear modulus. Since the fluid has a vanishing shear modulus, 
the traction can only be continuous if the appropriate strain components go to 
zero at these boundaries. However, due to the fact that 
\citet{2000MNRAS.319..902U} demanded that the crust be maximally strained at 
every point, the strain components have finite values at the interfaces and, 
therefore, one cannot ensure continuity of the traction. 

In defence of the \citet{2000MNRAS.319..902U} approach, one might argue that the 
shear modulus may be assumed to smoothly go to zero at the phase transitions. 
However, this is still problematic. As we show in Section~\ref{sec:Interface}, 
such 
an assumption means that one does not have enough equations to uniquely 
determine the displacement, in the case where one does not know the strain. A 
more realistic assumption might be to take almost the entire crust to be at 
breaking strain, with the exception of an infinitesimally small region at the 
boundaries where the displacement is adjusted to satisfy the continuity of the 
traction. 

Ultimately, the estimate (\ref{eq:Q_22UCB}) may give us an idea of the likely 
maximum mountain, but the calculation is not completely consistent. 

\subsection{Haskell, Jones \& Andersson}

\citet{2006MNRAS.373.1423H} set out to relax some of the assumptions made by 
\citet{2000MNRAS.319..902U}. This included dropping the Cowling approximation 
and ensuring the traction is continuous at the appropriate boundaries. They also 
noted that, by insisting the star is strained to the maximum throughout the 
crust, one loses the freedom to impose the boundary conditions of the problem. 

\citet{2006MNRAS.373.1423H} derived a system of coupled ordinary differential 
equations which describe the perturbations in the elastic crust and the fluid 
core relative to a spherically symmetric background star. They numerically 
integrated the perturbation equations and fixed the perturbation amplitude to 
the maximum necessary to begin to break the crust at a point, according the 
von Mises criterion. In their study, \citet{2006MNRAS.373.1423H} obtained the 
largest mountain when they assumed the core to be unperturbed, thus, allowing 
them to use a fully relativistic core combined with Newtonian perturbations in 
the crust. They reported a maximum quadrupole for a star with 
$M = \SI{1.4}{\solarMass}$, $R = \SI{10}{\kilo\metre}$ of 
\begin{equation}
    Q_{2 2}^\text{max} \approx \num{3.1e40} \ 
        \left( \frac{\bar{\sigma}_\text{max}}{10^{-1}} \right) \ 
        \si{\gram\centi\metre\squared}, 
\label{eq:Q_22HJA}
\end{equation}
which corresponds to an ellipticity of 
$\epsilon^\text{max} \approx \num{4.0e-5} \ 
(\bar{\sigma}_\text{max} / 10^{-1})$. This result is approximately an order of 
magnitude above that of \citet{2000MNRAS.319..902U}. 

The calculation of \citet{2006MNRAS.373.1423H} correctly treated the boundary 
condition at the crust-core interface by demanding that the traction was 
continuous. However, their calculation assumed the relaxed shape -- the strain 
is 
taken with respect to -- to be spherical. In general, the relaxed shape must be 
non-spherical, to give an equilibrium solution with a non-zero mountain. They 
did however stipulate that the surface shape of the star was deformed in an 
$(l, m) = (2,2)$ way. This effectively meant using an outer boundary condition 
where a traction-like force (i.e., a force per unit area) acts at the 
very surface of the star. Because of this, the maximum quadrupoles calculated in 
this framework turn out to be insensitive to the shear modulus of the crust, as 
they are sustained by this applied surface force. The lack of inclusion of a 
body force (i.e., a force per unit volume) in building the mountain 
meant that their formalism did not have the necessary freedom to ensure that the 
perturbed potential in the interior matches to the exterior solution. We discuss 
this particular subtlety further in Section~\ref{sec:Fluid}. 

By comparing with our new analysis we also note a number of typographical errors 
in their elastic perturbation equations. These errors turn out to have a 
surprisingly dramatic effect. Once they are corrected the maximum quadrupole 
increases by  three orders of magnitude, in sharp contrast with other estimates. 
This, in turn, highlights the conceptual problem with the formulation.

\subsection{Johnson-McDaniel \& Owen}

The most recent estimates for the largest possible mountain on a neutron star 
were provided by \citet{2013PhRvD..88d4004J}. They generalised the 
\citet{2000MNRAS.319..902U} argument to relativistic gravity while 
relaxing the Cowling approximation. They evaluated the required integral by 
employing a Green's function. For a $\SI{1.4}{\solarMass}$ star, described by 
the SLy equation of state \citep{2001A&A...380..151D}, they obtained the result, 
\begin{equation}
    Q_{2 2}^\text{max} \approx \num{2e39} \ 
        \left( \frac{\bar{\sigma}_\text{max}}{10^{-1}} \right) \ 
        \si{\gram\centi\metre\squared}, 
\label{eq:Q_22JMO}
\end{equation}
corresponding to $\epsilon^\text{max} \approx \num{3e-6} \ 
(\bar{\sigma}_\text{max} / 10^{-1})$. 

In following the \citet{2000MNRAS.319..902U} approach, the crust was taken to be 
strained to the maximum at every point, which means that the traction vector 
cannot be continuous at the crust boundaries. Furthermore,  they do not use the 
correct expression for the perturbed stress-energy tensor, since it does not 
include variations of the four-velocity. This may be a minor detail, but it 
should still be noted. 

In summary, although some of the above points may have a negligible impact on 
the maximum quadrupole estimates, there are issues with all previous studies 
of the maximum-mountain problem. 

\section{Building mountains}
\label{sec:Mountains}

In this section, we examine what must go into a consistent mountain calculation 
and discuss two methods for modelling mountains on neutron stars. The first 
approach, introduced in \citet{2000MNRAS.319..902U}, involves specifying the 
strain field associated with the mountain. We present a second method which, 
instead of starting with the strain, starts with a description of the perturbing 
force. Both approaches are valid and we demonstrate how they are equivalent. 

To help develop intuition, we will start by briefly discussing the case of 
strains built up in a spinning down star. We will therefore be considering the 
case of $(l, m) = (2, 0)$ perturbations relevant for rotational deformations, 
not the $(l, m) = (2, 2)$ relevant to the mountain case. Suppose a young 
neutron star with a molten crust spins  at an angular frequency, $\Omega$. At 
this rotation rate, the star cools and the crust solidifies. The star then 
begins to spin down to frequency, $\tilde{\Omega} < \Omega$.%
\footnote{This spin-down could be due to the usual radio emission that pulsars 
are well known for.}
Because the star has spun down, it changes shape according to the difference in 
the centrifugal force, $\propto (\Omega^2 - \tilde{\Omega}^2)$. This builds up 
strain in the crust as the shear stresses resist the change in shape. Should the 
star spin down sufficiently, the crust may fracture as stresses get too large. 
In fact, it has been suggested that the elastic yield of the crust in this 
process may be associated with the glitch phenomenon observed in some rotating 
pulsars \citep{1971AnPhy..66..816B, 2015MNRAS.446..865K}. 

Motivated by this example, which does not represent a neutron star mountain, we 
consider neutron star models forced away from sphericity by a perturbing force 
$f_i$, which we will choose to give mountain-like $(l, m) = (2, 2)$ 
perturbations. The elastic Euler equation (\ref{eq:PerturbedEulerElastic}) then 
becomes 
\begin{equation}
    0 = - \nabla_i p - \rho \nabla_i \Phi + \nabla^j t_{i j} + f_i ,
\label{eq:EulerElasticSource}
\end{equation}
We regard this equation as exact, and will consider perturbations of it below. 
In the fluid regions of the star, which cannot support shear stresses, the shear 
modulus goes to zero so the shear-stress tensor vanishes. To condense the 
notation, we define 
\begin{equation}
    H_i \equiv \nabla_i p + \rho \nabla_i \Phi, 
\end{equation}
which captures the familiar equation of hydrostatic equilibrium when $H_i = 0$. 
Therefore, the Euler equation (\ref{eq:EulerElasticSource}) can be expressed as 
\begin{equation}
    H_i = f_i + \nabla^j t_{i j}. 
\end{equation}
By considering a variation of $H_i$, we may write 
\begin{equation}
    \delta H_i = \nabla_i \delta p + \delta \rho \nabla_i \Phi 
        + \rho \nabla_i \delta \Phi, 
\label{eq:PerturbedH_i}
\end{equation}
where the perturbed quantities will need to be carefully defined in what 
follows. 

\begin{figure}
	\includegraphics[width=\columnwidth]{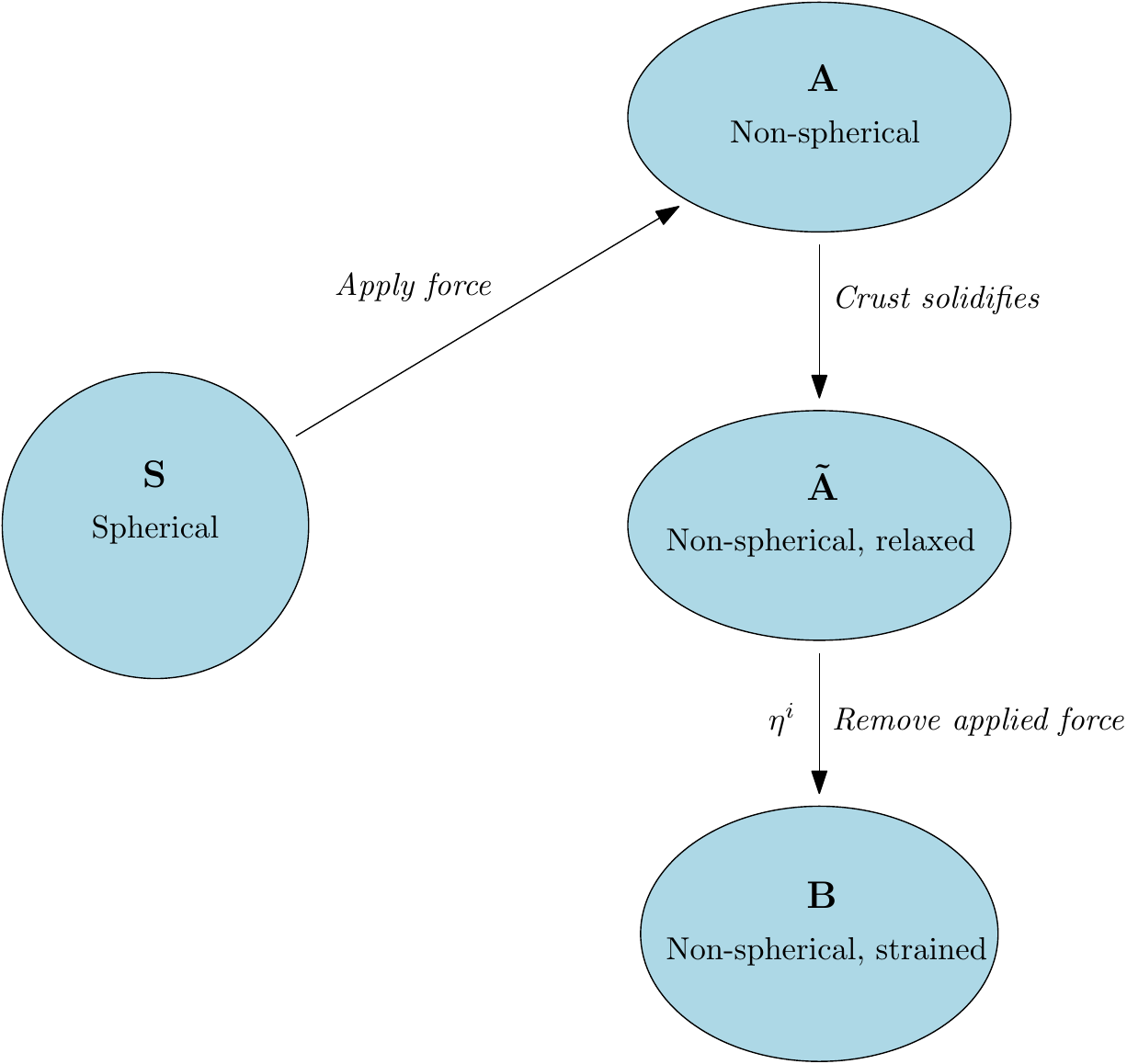}%
    \caption{\label{fig:StarsUCB} A schematic illustration showing the 
    configurations involved in mountain calculations. Note that previous 
    calculations have typically considered stars S and B, but not (explicitly) A 
    or \~{A}.}
\end{figure}

We now consider a family of four closely-related, equilibrium stars, 
illustrated in Fig.~\ref{fig:StarsUCB}: 

\begin{itemize}
    \item \textbf{Star S} -- A spherical, fluid star with 
        $(\rho_\text{S}, p_\text{S}, \Phi_\text{S})$: 
        \begin{equation}
            H_i^\text{S} = 0. 
        \label{eq:EulerS}
        \end{equation}
    \item \textbf{Star A} -- A force is applied to star S, which produces a 
        non-spherical, fluid star with 
        $(\rho_\text{A}, p_\text{A}, \Phi_\text{A})$: 
        \begin{equation}
            H_i^\text{A} = f_i. 
        \label{eq:EulerA}
        \end{equation}
    \item \textbf{Star \~{A}} -- The crust of star A solidifies while the force 
        is maintained. This gives rise to a non-spherical, relaxed star with the 
        same structure as star A, although (formally) with a non-zero shear 
        modulus. The star has $(\rho_\text{\~{A}} = \rho_\text{A}, 
        p_\text{\~{A}} = p_\text{A}, \Phi_\text{\~{A}} = \Phi_\text{A})$: 
        \begin{equation}
            H_i^\text{\~{A}} = H_i^\text{A} = f_i. 
        \end{equation}
        Note that, because star A and star \~{A} have the same shape, in 
        general, we only need to refer to star A in the following discussion 
        when specifying the values of perturbed quantities. 
    \item \textbf{Star B} -- The force on star \~{A} is removed, which builds up 
        strain in the crust. The associated deformation between these two stars 
        is described by the Lagrangian displacement vector field, $\eta^i$. The 
        star is non-spherical and strained with 
        $(\rho_\text{B}, p_\text{B}, \Phi_\text{B})$: 
        \begin{equation}
            H_i^\text{B} = \nabla^j t_{i j}(\eta). 
        \label{eq:EulerB}
        \end{equation}
        Note that it is this star, star B, that we are ultimately interested in: 
        this is the star with a mountain supported in a self-consistent way by 
        elastic strains, with no external force acting.
\end{itemize}

Note that the force $f_i$ has a simple physical interpretation: it is the force 
that, when applied to our equilibrium star with the mountain (star B), takes us 
to the corresponding unstrained star (star A, or equivalently \~{A}). Note, 
however, that there is no requirement whatsoever that, in the real world, this 
force ever acted upon our star. For a realistic situation, the elastic strains 
that support the deformation of star B will likely have evolved through some 
complex process of plastic flow and cracking, possibly combined with whatever 
agent causes the asymmetry to develop. The usefulness of $f_i$ is two-fold. 
Firstly, it allows us to explicitly identify the unstrained configuration. 
Secondly, the explicit introduction of the force into the Euler equation 
provides the necessary freedom to determine the displacement vector and satisfy 
all the boundary conditions. 

It is useful to consider the differences between the stellar models described 
above. Thus, we introduce the notation, 
\begin{equation}
    \delta H_i^\text{AB} = H_i^\text{B} - H_i^\text{A}, 
\end{equation}
i.e., $\delta H_i^\text{AB}$ is the quantity that must be added to 
$H_i^\text{A}$ to obtain $H_i^\text{B}$. 

The difference between star B (\ref{eq:EulerB}) and star S (\ref{eq:EulerS}) is 
\begin{equation}
    \delta H_i^\text{SB} = \nabla^j t_{i j}(\eta). 
\label{eq:PerturbedEulerUCB}
\end{equation}
Expression (\ref{eq:PerturbedEulerUCB}) relates perturbations between the 
strained star -- with a mountain -- and the spherical, reference star to the 
shear 
stresses induced when the relaxed star is deformed according to the 
displacement, $\eta^i$. This is the standard picture for understanding neutron 
star mountains and, indeed, it is this expression that is used to estimate the 
maximum quadrupole in \citet{2000MNRAS.319..902U} and 
\citet{2013PhRvD..88d4004J}. It is 
important to note that in these calculations one does not have to determine the 
relaxed shape and, indeed, stars A and \~{A} did not appear explicitly in 
previous calculations. However, we demonstrate that the relaxed shape is, in 
principle, calculable in Appendix~\ref{app:Appendix}. 

\begin{figure}
	\includegraphics[width=\columnwidth]{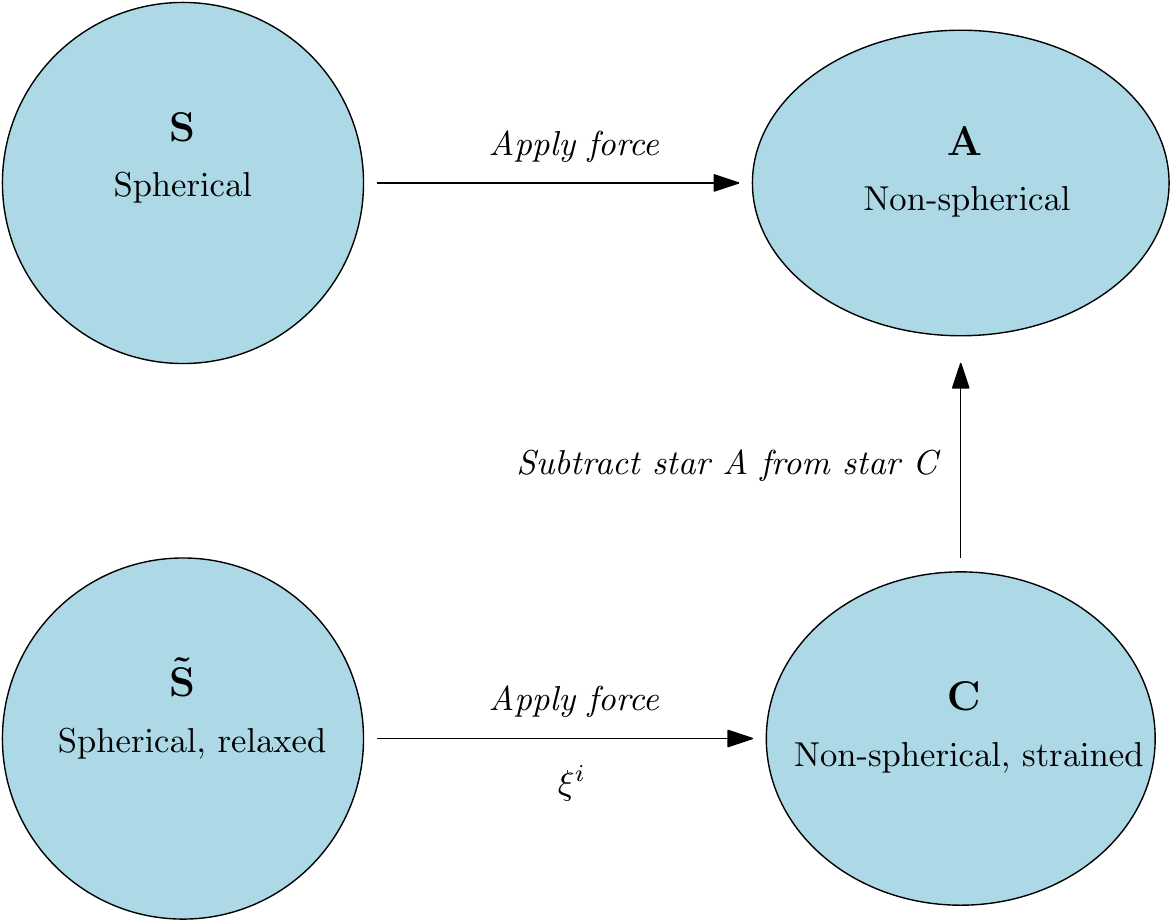}%
    \caption{\label{fig:StarsScheme}A schematic illustration showing the 
    configurations in the force-based mountain scheme.}
\end{figure}

As we discuss in more detail below, for a fully consistent calculation that 
satisfies all the boundary conditions of the problem it is not convenient to 
use (\ref{eq:PerturbedEulerUCB}) alone. Rather, we present an alternative 
strategy which makes explicit use of the deforming force. To this end, we 
introduce two additional stars shown in Fig.~\ref{fig:StarsScheme}: 

\begin{itemize}
    \item \textbf{Star \~{S}} -- The crust of star S solidifies. This star has 
        the same shape as star S with a non-zero shear modulus and 
        $(\rho_\text{\~{S}} = \rho_\text{S}, p_\text{\~{S}} = p_\text{S}, 
        \Phi_\text{\~{S}} = \Phi_\text{S})$: 
        \begin{equation}
            H_i^\text{\~{S}} = H_i^\text{S} = 0. 
        \end{equation}
    \item \textbf{Star C} -- A force is applied to star \~{S}. This induces 
        stress in the crust, described by the Lagrangian displacement, $\xi^i$, 
        and produces a non-spherical, strained star with 
        $(\rho_\text{C}, p_\text{C}, \Phi_\text{C})$: 
        \begin{equation}
            H_i^\text{C} = f_i + \nabla^j t_{i j}(\xi). 
        \label{eq:EulerC}
        \end{equation}
\end{itemize}

We can then consider the difference between stars A and C. By using 
(\ref{eq:EulerA}) and (\ref{eq:EulerC}), we obtain 
\begin{equation}
    \delta H_i^\text{AC} = \nabla^j t_{i j}(\xi). 
\label{eq:PerturbedEulerScheme}
\end{equation}

We can note the similarity of (\ref{eq:PerturbedEulerScheme}) to 
(\ref{eq:PerturbedEulerUCB}). Indeed, comparing Figs.~\ref{fig:StarsUCB} 
and \ref{fig:StarsScheme}, we note the following. In Fig.~\ref{fig:StarsUCB}, 
the addition of force $f_i$ maps star B to star A, generating a displacement 
$-\eta^i$, while in Fig.~\ref{fig:StarsScheme}, the addition of the force $f_i$ 
maps star \~{S} to star C, generating a displacement field $\xi^i$. It follows 
that, to a good approximation, these vector fields are related by 
\begin{equation}
    \eta^i =  - \xi^i.
\end{equation}
Comparing (\ref{eq:PerturbedEulerScheme}) to (\ref{eq:PerturbedEulerUCB}) then 
gives the corresponding relation between the associated scalar perturbations, 
\begin{equation}
    \delta H_i^\text{SB} = - \delta H_i^\text{AC}.
\label{eq:SB_AC_relation}
\end{equation}
These relations are not exact, as in Fig.~\ref{fig:StarsUCB} the force $f_i$ 
acts upon star B, while in Fig.~\ref{fig:StarsScheme} it acts upon star \~{S}, 
but these two stars themselves differ from one another only in a perturbative 
way, so the difference in the action of $f_i$ on the two must be of second 
order.

This immediately suggests a strategy for computing the deformation of star B 
(i.e., $\delta H_i^\text{SB}$ and other perturbed quantities). We can 
easily compute the perturbations linking S and A (i.e., 
$\delta H_i^\text{SA}$ etc.), as this is just the perturbation of a 
spherical fluid star by the force $f_i$. We can, with only a little more effort, 
compute the perturbations linking star \~{S} and C (i.e., 
$\delta H_i^\text{SC}$ etc.), as this is just the perturbation of a 
spherical elastic star by $f_i$. Then, we can take the difference between these 
two configurations to give the difference between star A and C (i.e., 
$\delta H_i^\text{AC}$ etc.), which is, up to an overall sign, the 
deformation of star B relative to star S that we require, as per 
(\ref{eq:SB_AC_relation}).

As we are interested in computing maximum mountains, we will choose the force 
$f_i$ such that the breaking strain is reached at some point in the crust of 
star C. Note, however, that in this force-based approach, we will not be able 
to follow \citet{2000MNRAS.319..902U} and find the solution where the strain is 
reached at \textit{all} points (simultaneously) in the crust. This is a price 
one pays in adopting the force-based approach. We have, however, reduced the 
calculation of the mountain to two simpler calculations, both taking place on a 
spherical background and with readily implementable boundary conditions. We now 
go on to consider these calculations in detail.

\section{Perturbation formalism}
\label{sec:Perturbation}

In order to develop the second strategy in detail, we take the background star 
to be non-rotating and the perturbations to be static. Our star is separated 
into three layers: a fluid core, an elastic crust and a fluid ocean. We choose 
to include a fluid layer outside the crust since at low densities the crustal 
lattice begins to melt and it also simplifies the matching to the exterior 
gravitational potential \citep*{2020PhRvD.101j3025G}. The crust only comes into 
the structure equations at the perturbative level. 

From equations~(\ref{eq:Continuity}), (\ref{eq:Euler}) and (\ref{eq:Poissons}), 
we 
obtain the Newtonian equations of structure for a non-rotating, fluid star, 
\begin{subequations}
\begin{gather}
    m' = 4 \uppi r^2 \rho, \\
    p' = - \rho \Phi' 
\end{gather}
and 
\begin{equation}
    \Phi' = \frac{G m}{r^2}, 
\end{equation}
\end{subequations}
where $m(r)$ is the mass enclosed in radius $r$. These equations are closed by 
supplying an equation of state (\ref{eq:EOS}). 

\subsection{\label{sec:Fluid}Fluid perturbations}

In order to calculate the relaxed configuration (star A), we need to introduce 
the force $f_i$. For practical purposes, it is convenient to write the force as 
the gradient of a potential, $\chi$, 
\begin{equation}
    f_i = - \rho \nabla_i \chi. 
\label{eq:Force}
\end{equation}
This is not the most general expression for the force but it allows us to 
combine $\chi$ with the gravitational potential, which simplifies the analysis. 
To make the notation more compact, we introduce the total perturbed potential, 
$U = \delta \Phi + \chi$. 

At this point, we note that this is where our calculation differs from previous 
work \citep{2000MNRAS.319..902U, 2006MNRAS.373.1423H, 2013PhRvD..88d4004J}. 
Previous calculations set out to evaluate the perturbed Euler equation 
(\ref{eq:PerturbedEulerUCB}) where the strain is taken with respect to the 
relaxed shape the crust wants to have (see Fig.~\ref{fig:StarsUCB}). In our 
method, we start with the deforming force and evaluate 
(\ref{eq:PerturbedEulerScheme}), using the subtraction scheme (taking the 
difference between stars C and A) set out in Section~\ref{sec:Mountains}. The 
use of this force is a subtle, but important, detail since without it one does 
not have the necessary freedom to impose all the boundary conditions of the 
problem.  We emphasise this point since this issue was somewhat confused in the 
analysis of \citet{2006MNRAS.373.1423H} who calculate perturbations of a 
spherical star but do not explicitly consider the force which sources them. It 
is for this reason that they were unable to satisfy the boundary condition on 
the potential at the surface. This point is elucidated below. 

Recall that, as we discussed earlier, we assume all perturbed quantities to 
be expanded in spherical harmonics, but it will be sufficient for our 
discussion to focus on the $(l, m) = (2, 2)$ mode. The system of equations which 
describes fluid perturbations then simplifies to a single second-order 
differential equation for the perturbed potential. From the perturbed Poisson's 
equation (\ref{eq:PerturbedPoissons}), we get 
\begin{subequations}\label{eqs:FluidPerturbations}
\begin{equation}
    \delta \Phi'' + \frac{2}{r} \delta \Phi' - \frac{\beta^2}{r^2} \delta \Phi 
        = 4 \uppi G \delta \rho, 
\label{eq:PerturbedPoissonsFluid}
\end{equation}
where $\beta \equiv \sqrt{l (l + 1)}$. The perturbed Euler equation 
(\ref{eq:PerturbedEulerStatic}) returns 
\begin{equation}
    \delta \rho = - \frac{\rho}{c_\text{s}^2} U. 
\end{equation}
\end{subequations}
Therefore, provided a description of the perturbing force, 
equations~(\ref{eqs:FluidPerturbations}) give a second-order equation that 
describes 
the perturbations in the fluid. 

The perturbed potential must satisfy two boundary conditions. At the centre of 
the star the solution must be regular and at the surface it must match to the 
external solution. Therefore, in addition to $\chi$ being regular at the centre 
of the star and continuous at all interfaces, we must have 
\begin{subequations}\label{eqs:PerturbedPotentialBoundaries}
\begin{equation}
    \delta \Phi(0) = 0 
\label{eq:PerturbedPotentialCentre}
\end{equation}
and 
\begin{equation}
    R \delta \Phi'(R) = - (l + 1) \delta \Phi(R). 
\label{eq:PerturbedPotentialSurface}
\end{equation}
\end{subequations}
From regularity, we obtain an initial condition by expanding 
(\ref{eqs:FluidPerturbations}) in small $r$, 
\begin{equation}
    \delta \Phi(r) = a_0 r^l [1 + \mathcal{O}(r^2)], 
\end{equation}
where $a_0$ is a constant which parametrises the amplitude of the perturbations. 
In the case when $\chi = 0$ and there is no driving force, this initial 
condition provides sufficient information to calculate the perturbations up to 
the surface. At the surface, however, there is no freedom left to impose the 
surface boundary condition (\ref{eq:PerturbedPotentialSurface}) -- except in the 
special case of $a_0 = 0$ where there are no perturbations. [This is the issue 
that the formalism of \citet{2006MNRAS.373.1423H} suffers from, and why in that 
analysis a surface force had to be effectively introduced via a boundary 
condition.] This serves as a simple demonstration of the fact that an unforced, 
fluid equilibrium is a spherical star. Equations~(\ref{eqs:FluidPerturbations}) 
with the boundary conditions (\ref{eqs:PerturbedPotentialBoundaries}) provide 
the necessary information to calculate perturbations in the fluid regions of the 
star sourced by a perturbing force. 

\subsection{The crust}

In order to calculate the strained star (star C) in our scheme outlined in 
Section~\ref{sec:Mountains} (Fig.~\ref{fig:StarsScheme}), we must consider the 
role 
of the elastic crust. We reiterate that we consider perturbations with respect 
to a spherical, reference star. 

The elastic material is characterised by the shear-stress tensor, 
\begin{equation}
    t_{i j} = \mu \left( \nabla_i \xi_j + \nabla_j \xi_i 
        - \frac{2}{3} g_{i j} \nabla_k \xi^k \right), 
\end{equation}
where $\mu$ is the shear modulus of the crust and $g_{i j}$ is the flat 
three-metric. Note that we also have 
\begin{equation}
    t_{i j} = 2 \mu \sigma_{i j}, 
\end{equation}
where $\sigma_{i j}$ is the stress tensor. [This is a factor of two different 
to the expressions in \citet{2000MNRAS.319..902U} and 
\citet{2006MNRAS.373.1423H} but the 
same as used in \citet{2013PhRvD..88d4004J}.] We use the static displacement 
vector appropriate for polar perturbations \citep{2000MNRAS.319..902U}, 
\begin{equation}
    \xi^i = \xi_r(r) \nabla^i r Y_{l m} 
        + \frac{r}{\beta} \xi_\perp(r) \nabla^i Y_{l m}. 
\end{equation}

To make the application of the boundary conditions straightforward, we consider 
the perturbed traction vector, which may be identified from the perturbed Euler 
equation (\ref{eq:PerturbedEulerElastic}), 
\begin{equation}
\begin{split}
    T^i &= (\delta p g^{i j} - t^{i j}) \nabla_j r \\
        &= [\delta p(r) - T_1(r)] \nabla^i r Y_{l m} 
        - r T_2(r) \nabla^i Y_{l m}, 
\end{split}
\label{eq:PerturbedTraction}
\end{equation}
where we have defined the following two variables related to the radial and 
tangential components of the traction: 
\begin{subequations}\label{eqs:TractionVariables}
\begin{equation}
    T_1(r) Y_{l m} \equiv t_{r r} = \frac{2 \mu}{3 r} (- 2 \xi_r 
        + \beta \xi_\perp + 2 r \xi_r') Y_{l m} 
\end{equation}
and 
\begin{equation}
    T_2(r) \nabla_\theta Y_{l m} \equiv \frac{t_{r \theta}}{r} 
        = \frac{\mu}{\beta r} (\beta \xi_r - \xi_\perp + r \xi_\perp') 
        \nabla_\theta Y_{l m}. 
\label{eq:T_2}
\end{equation}
\end{subequations}

From the perturbed continuity equation (\ref{eq:PerturbedContinuity}), we then 
obtain 
\begin{equation}
\begin{split}
    \delta \rho &= - \rho \xi_r' - \left( \frac{2 \rho}{r} + \rho' \right) \xi_r 
        + \frac{\beta \rho}{r} \xi_\perp \\
        &= - \left( \frac{3 \rho}{r} + \rho' \right) \xi_r 
        + \frac{3 \beta \rho}{2 r} \xi_\perp - \frac{3 \rho}{4 \mu} T_1.
\end{split}
\label{eq:PerturbedContinuityElastic} 
\end{equation}

From the definitions of the traction variables (\ref{eqs:TractionVariables}), we 
have the following differential equations which describe the displacement 
vector: 
\begin{subequations}\label{eqs:ElasticPerturbations}
\begin{equation}
    \xi_r' = \frac{1}{r} \xi_r - \frac{\beta}{2 r} \xi_\perp 
        + \frac{3}{4 \mu} T_1 
\end{equation}
and 
\begin{equation}
    \xi_\perp' = - \frac{\beta}{r} \xi_r + \frac{1}{r} \xi_\perp 
        + \frac{\beta}{\mu} T_2. 
\end{equation}
From the radial part of the perturbed Euler equation 
(\ref{eq:PerturbedEulerElastic}) combined with the perturbed continuity equation 
(\ref{eq:PerturbedContinuityElastic}), 
\begin{equation}
\begin{split}
    \bigg( 1 + &\frac{3 c_\text{s}^2 \rho}{4 \mu} \bigg) T_1' 
        = \rho U' \\
        - &\left[ (c_\text{s}^2)' (3 \rho + r \rho') 
        + c_\text{s}^2 \left( \frac{3 \beta^2 \rho}{2 r} + \rho' 
        - \frac{r \rho'^2}{\rho} 
        + r \rho'' \right) \right] \frac{1}{r} \xi_r \\
        + &\left[ (c_\text{s}^2)' 3 \rho + c_\text{s}^2 
        \left( \frac{3 \rho}{r} + \rho' \right) \right] 
        \frac{\beta}{2 r} \xi_\perp \\
        - &\left[ \frac{3}{r} 
        + (c_\text{s}^2)' \frac{3 \rho}{4 \mu} 
        + c_\text{s}^2 \left( \frac{3 \rho}{r} - \frac{\rho \mu'}{\mu} 
        + \rho' \right) \frac{3}{4 \mu} \right] T_1 \\
        + &\left( 1 + \frac{3 c_\text{s}^2 \rho}{2 \mu}\right) 
        \frac{\beta^2}{r} T_2. 
\end{split}
\end{equation}
Then, from the tangential piece of (\ref{eq:PerturbedEulerElastic}) we find 
\begin{equation}
\begin{split}
    T_2' = \frac{\rho}{r} U 
        - &c_\text{s}^2 (3 \rho + r \rho') \frac{1}{r^2} \xi_r \\
        + &\left[ \frac{3 c_\text{s}^2 \rho}{2} 
        + \left( 1 - \frac{2}{\beta^2} \right) \mu \right] \frac{\beta}{r^2} 
        \xi_\perp \\
        + &\left( \frac{1}{2} - \frac{3 c_\text{s}^2 \rho}{4 \mu} \right) 
        \frac{1}{r} T_1 - \frac{3}{r} T_2. 
\end{split}
\end{equation}
We also have the perturbed Poisson's equation (\ref{eq:PerturbedPoissonsFluid}), 
which combines with the perturbed continuity equation 
(\ref{eq:PerturbedContinuityElastic}) to give 
\begin{equation}
\begin{split}
    \delta \Phi'' + \frac{2}{r} \delta \Phi' - \frac{\beta^2}{r^2} \delta \Phi 
        = - &4 \uppi G \left( \frac{3 \rho}{r} + \rho' \right) \xi_r \\
        + &6 \uppi G \frac{\beta \rho}{r} \xi_\perp 
        - 3 \uppi G \frac{\rho}{\mu} T_1.
\end{split}
\end{equation}
\end{subequations}
Equations~(\ref{eqs:ElasticPerturbations}) form a coupled system of ordinary 
differential equations to describe the perturbations in the elastic material. 
We have compared our perturbation equations with that of 
\citet{2006MNRAS.373.1423H} (in the limit of $\chi = 0$) and noted several 
discrepancies. We find that these mistakes increase the maximum quadrupole 
estimates of \citet{2006MNRAS.373.1423H} by three orders of magnitude. 

\subsection{\label{sec:Interface}Interface conditions}

At this point, we address the boundary conditions at the fluid-elastic 
interfaces since we wish to connect perturbations in the fluid core and ocean 
with the elastic crust. Provided the density is smooth (which we assume), the 
perturbed potential, $\delta \Phi$, and its derivative, $\delta \Phi'$, must be 
continuous at an interface. To see how the other perturbed quantities behave at 
an interface, we must consider the perturbed traction 
(\ref{eq:PerturbedTraction}). 

This admits two quantities which must be continuous: the radial and tangential 
components. Since the shear modulus vanishes in the fluid, continuity of the 
radial traction, $(\delta p - T_1)$, provides an algebraic relation which must 
hold true at an interface, 
\begin{equation}
\begin{split}
    \rho U_\text{F} = 
        &\left( 1 + \frac{3 c_\text{s}^2 \rho}{4 \mu} \right) T_{1 \text{E}} \\
        &+ c_\text{s}^2 \left[ \left( \frac{3 \rho}{r} 
        + \rho' \right) \xi_{r \text{E}} 
        - \frac{3 \beta \rho}{2 r} \xi_{\perp \text{E}} \right], 
\end{split}
\label{eq:RadialTraction}
\end{equation}
where the subscripts F and E denote the fluid and elastic sides of the 
interface, respectively. We note that the radial displacement, $\xi_r$ must be 
continuous at a boundary, however, this does not necessarily have to be the case 
for the tangential piece, $\xi_\perp$. From the tangential part of the traction, 
we have $T_2 = 0$ at a fluid-elastic interface. 

In reference to the maximally strained approach of 
\citet{2000MNRAS.319..902U} and \citet{2013PhRvD..88d4004J}, we note that, if 
one assumes 
the shear modulus smoothly goes to zero at a fluid-elastic interface, then the 
tangential traction condition is trivially satisfied [see (\ref{eq:T_2})]. This 
would effectively result in the displacement vector in the crust being arbitrary 
since there are not enough boundary conditions to constrain it. It is not clear 
how to resolve this issue. 

In the fluid regions of the star, the perturbations are governed by 
equations~(\ref{eqs:FluidPerturbations}) and so are described by the variables, 
$(\delta \Phi', \delta \Phi)$. In the crust, we have a more complex structure 
with equations~(\ref{eqs:ElasticPerturbations}) and quantities, 
$(\delta \Phi', \delta \Phi, \xi_r, \xi_\perp, T_1, T_2)$. We assume the force 
is known. The perturbations in the elastic crust present a boundary-value 
problem. For the six variables, we have six boundary conditions: continuity of 
$\delta \Phi'$ and $\delta \Phi$ at the core-crust transition and the two 
traction conditions -- (\ref{eq:RadialTraction}) and $T_2 = 0$ -- at both 
interfaces. Therefore, the problem is well posed. 

Additionally, it is straightforward to show that the boundary condition on the 
Lagrangian variation of the pressure, $\Delta p(R) = 0$, is trivially 
satisfied by the background structure. 

\section{The deforming force}
\label{sec:Sources}

The formalism we detail above requires a description of the deforming force 
which causes the star to have a non-spherical shape. Because of the abstract 
nature of this force, it is difficult to prescribe without a detailed 
evolutionary calculation of the history of the star. As a proof-of-principle 
calculation, we examine three example sources. 

We use a polytropic equation of state, 
\begin{equation}
    p(\rho) = K \rho^{1 + 1/n}, 
\end{equation}
where $K$ is a constant of proportionality and $n$ is the polytropic constant. 
We work with $n = 1$ and generate background models with 
$M = \SI{1.4}{\solarMass}$, $R = \SI{10}{\kilo\metre}$. For the shear-modulus 
profile in the crust, we consider a simple linear model 
\citep{2006MNRAS.373.1423H}, 
\begin{equation}
    \mu(\rho) = \kappa \rho, 
\end{equation}
where $\kappa = \SI{e16}{\centi\metre\squared\per\second\squared}$. We assume 
the core-crust transition to occur at 
$\rho_\text{base} = \SI{2e14}{\gram\per\centi\metre\cubed}$ [which is the same 
as \citet{2000MNRAS.319..902U}], while the crust-ocean transition is at 
$\rho_\text{top} = \SI{e6}{\gram\per\centi\metre\cubed}$ 
\citep{2020PhRvD.101j3025G}. 

We consider three sources for the perturbations: (i) a potential which satisfies 
Laplace's equation, (ii) a potential which satisfies Laplace's equation but does 
not act in the core and (iii) a thermal pressure perturbation. For each 
prescription, we generate two stars -- a relaxed star, which experiences purely 
fluid perturbations (star A in Fig.~\ref{fig:StarsScheme}), and a strained star, 
which experiences elastic perturbations in the crust (star C in 
Fig.~\ref{fig:StarsScheme}). We normalise the perturbations by ensuring the 
strained star reaches breaking strain at a point in the crust, subject to the 
von Mises criterion, and that the relaxed star experiences the same force. This 
allows us to work out the quadrupole moment of each star. Our results for the 
three sources are summarised in Table~\ref{tab:ellipticities}. 

\begin{table*}
    \caption{\label{tab:ellipticities}The maximum quadrupoles and ellipticities 
             from the different models. For each case, we show the quadrupole, 
             $Q_{2 2}^\text{relaxed}$, and ellipticity, 
             $\epsilon^\text{relaxed}$, for the relaxed star and the difference 
             relative to the strained star with quadrupole 
             $Q_{2 2}^\text{strained}$ and ellipticity 
             $\epsilon^\text{strained}$.}
    \begin{tabular}{ l c c c c }
        \hline\hline
        Source & $|Q_{2 2}^\text{relaxed}|$ / \si{\gram\centi\metre\squared} 
        & $|\epsilon^\text{relaxed}|$ 
        & $|Q_{2 2}^\text{strained} - Q_{2 2}^\text{relaxed}|$ 
        / \si{\gram\centi\metre\squared} 
        & $|\epsilon^\text{strained} - \epsilon^\text{relaxed}|$ \\ 
        \hline
        Solution of Laplace's equation & \num{2.4e43} & \num{3.1e-2} 
        & \num{1.7e37} & \num{2.2e-8} \\ 
        Solution of Laplace's equation (outside core) & \num{1.4e41} 
        & \num{1.8e-4} & \num{4.4e38} & \num{5.7e-7} \\ 
        Thermal pressure perturbation & \num{9.2e38} & \num{1.2e-6} 
        & \num{4.0e38} & \num{5.2e-7} \\
        \hline
    \end{tabular}
\end{table*}

To solve the coupled sets of ordinary differential 
equations~(\ref{eqs:FluidPerturbations}) and (\ref{eqs:ElasticPerturbations}), 
we used an explicit fifth-order Runge-Kutta method implemented in the  
\texttt{solve\_ivp} function from the \textsc{scipy} library. The structure in 
the fluid core may be straightforwardly calculated using 
equations~(\ref{eqs:FluidPerturbations}) with boundary 
condition~(\ref{eq:PerturbedPotentialCentre}). The crust presents a 
boundary-value problem with the interface conditions described in 
Section~\ref{sec:Interface}. There are a variety of numerical techniques to 
solve such problems. Due to the linearity of the 
system~(\ref{eqs:ElasticPerturbations}), we used the linearly independent scheme 
described in Appendix~B of \citet{2020PhRvD.101j3025G}. With the perturbations 
in the crust, one can integrate equations~(\ref{eqs:FluidPerturbations}) through 
the fluid ocean to the surface. At this point, one can verify that the boundary 
condition at the surface (\ref{eq:PerturbedPotentialSurface}) is automatically 
satisfied.

\subsection{A solution of Laplace's equation}

The first example we consider is based on the form of the deforming 
potential for tidal deformations [see, e.g., \citet{2020PhRvD.101h3001A}]. The 
source potential is taken to be a solution of Laplace's equation, 
\begin{equation}
    \nabla^2 \chi = 0. 
\label{eq:Laplaces}
\end{equation}
This example is particularly convenient since the perturbed Poisson's equation 
(\ref{eq:PerturbedPoissons}) is simply modified by $\delta \Phi \rightarrow U$. 
Therefore, we may write 
\begin{equation}
    \nabla^2 U = 4 \uppi G \delta \rho. 
\end{equation}
The total perturbed potential must be regular at the origin, $U(0) = 0$. 

By using the definition of the multipole moment (\ref{eq:Multipole}) with the 
perturbed Poisson's equation (\ref{eq:PerturbedPoissonsFluid}), one can show 
through integration by parts, 
\begin{equation}
    Q_{l m} = - \frac{(2 l + 1) R^{l + 1}}{4 \uppi G} \delta \Phi(R), 
\label{eq:Multipole2}
\end{equation}
where the boundary conditions~(\ref{eqs:PerturbedPotentialBoundaries}) have been 
used for simplification. This result, perhaps more familiar in relativistic 
calculations, shows that one can obtain the multipole moments from the 
variations of the potential at the surface. One can also write the multipole in 
terms of the total perturbed potential, 
\begin{equation}
    Q_{l m} = \frac{R^{l + 1}}{4 \uppi G} [R U'(R) - l U(R)]. 
\label{eq:Multipole3}
\end{equation}
The advantage of writing the multipole in this way is that one does not need to 
disentangle the two potentials ($\chi$ and $\delta \Phi$)  from $U$. 

The source potential must be regular at the centre and so it must be of the 
form, 
\begin{equation}
    \chi(r) = A r^l, 
\end{equation}
where $A$ is a constant. Its value will be chosen to ensure the star is 
maximally strained at some point in the crust. The source potential at the 
surface is given by 
\begin{equation}
    \chi(R) = \frac{1}{2 l + 1} [R U'(R) + (l + 1) U(R)]. 
\label{eq:SourcePotentialSurface}
\end{equation}
It is this quantity that we use to ensure that the relaxed and strained stars 
experience the same force. 

To make sure the star is maximally strained we calculate the von Mises 
strain, $\bar{\sigma}$. The von Mises strain is defined using the strain tensor, 
\begin{equation}
    \bar{\sigma}^2 \equiv \frac{1}{2} \sigma_{i j} \sigma^{i j}. 
\end{equation}
The von Mises criterion states that an elastic material will reach its yield 
limit when $\bar{\sigma} \geq \bar{\sigma}_\text{max}$. For $(l, m) = (2, 2)$ 
perturbations, we have 
\begin{equation}
\begin{split}
    \bar{\sigma}^2 = \frac{5}{256 \uppi} \Bigg\{ 6 \sin^2 \theta 
        \Bigg[ 3 \sin^2 \theta \cos^2 2 \phi &\left( \frac{T_1}{\mu} \right)^2 \\
        + 4 (3 + \cos 2 \theta - 2 \sin^2 \theta \cos 4 \phi) 
        &\left( \frac{T_2}{\mu} \right)^2 \Bigg] \\
        + (35 + 28 \cos 2 \theta + \cos 4 \theta + 8 \sin^4 \theta \cos 4 \phi) 
        &\left( \frac{\xi_\perp}{r} \right)^2 \Bigg\}. 
\end{split}
\label{eq:vonMises}
\end{equation}
Since the von Mises strain is a function of position, we can identify where the 
strain is highest (and, thus, the crust will break first) and take that point to 
be at breaking strain, which we assume to be $\bar{\sigma}_\text{max} = 10^{-1}$ 
\citep{2009PhRvL.102s1102H}. 

\begin{figure*}
	\includegraphics[width=0.49\textwidth]{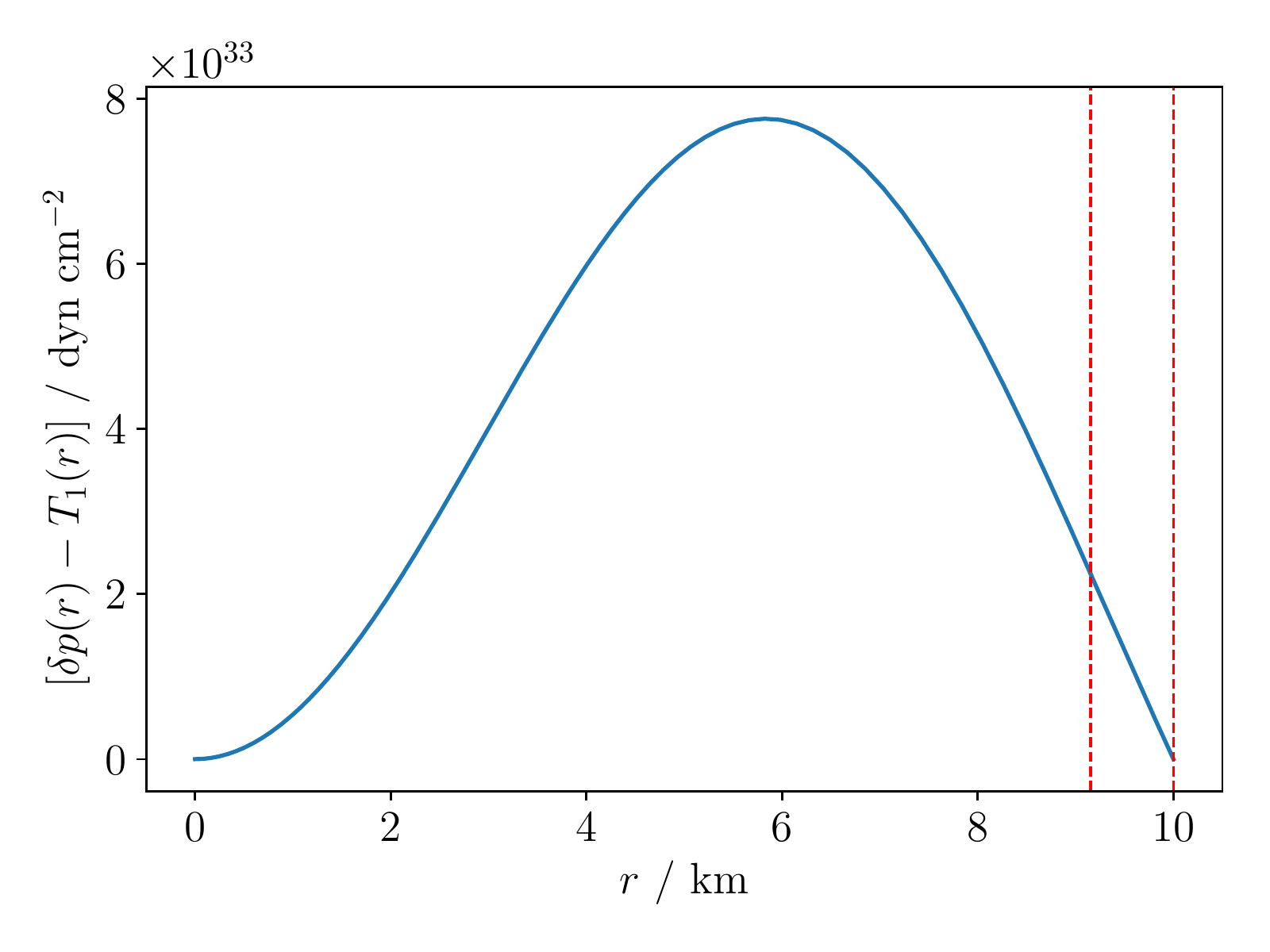}%
    \includegraphics[width=0.49\textwidth]{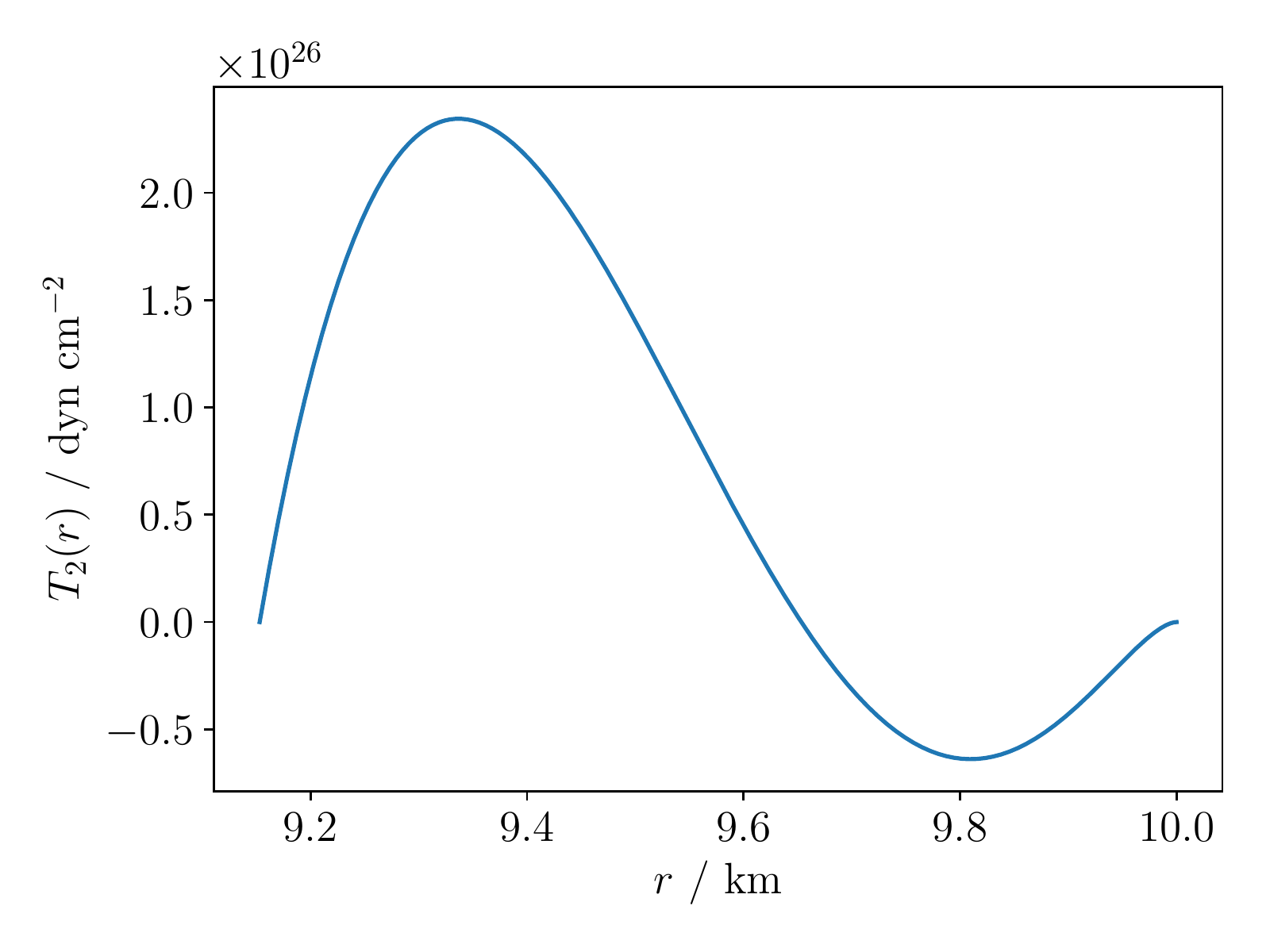}%
    \caption{\label{fig:Traction}The radial (left panel) and tangential (right 
             panel) components of the perturbed traction as functions of radius 
             for the potential solution to Laplace's equation. 
             The vertical red dashed lines in the left panel indicate the base 
             and the top of the crust. Regarding the horizontal range in the 
             right panel, recall that $T_2$ only has a finite value in the 
             crust.}
\end{figure*}

\begin{figure}
	\includegraphics[width=\columnwidth]{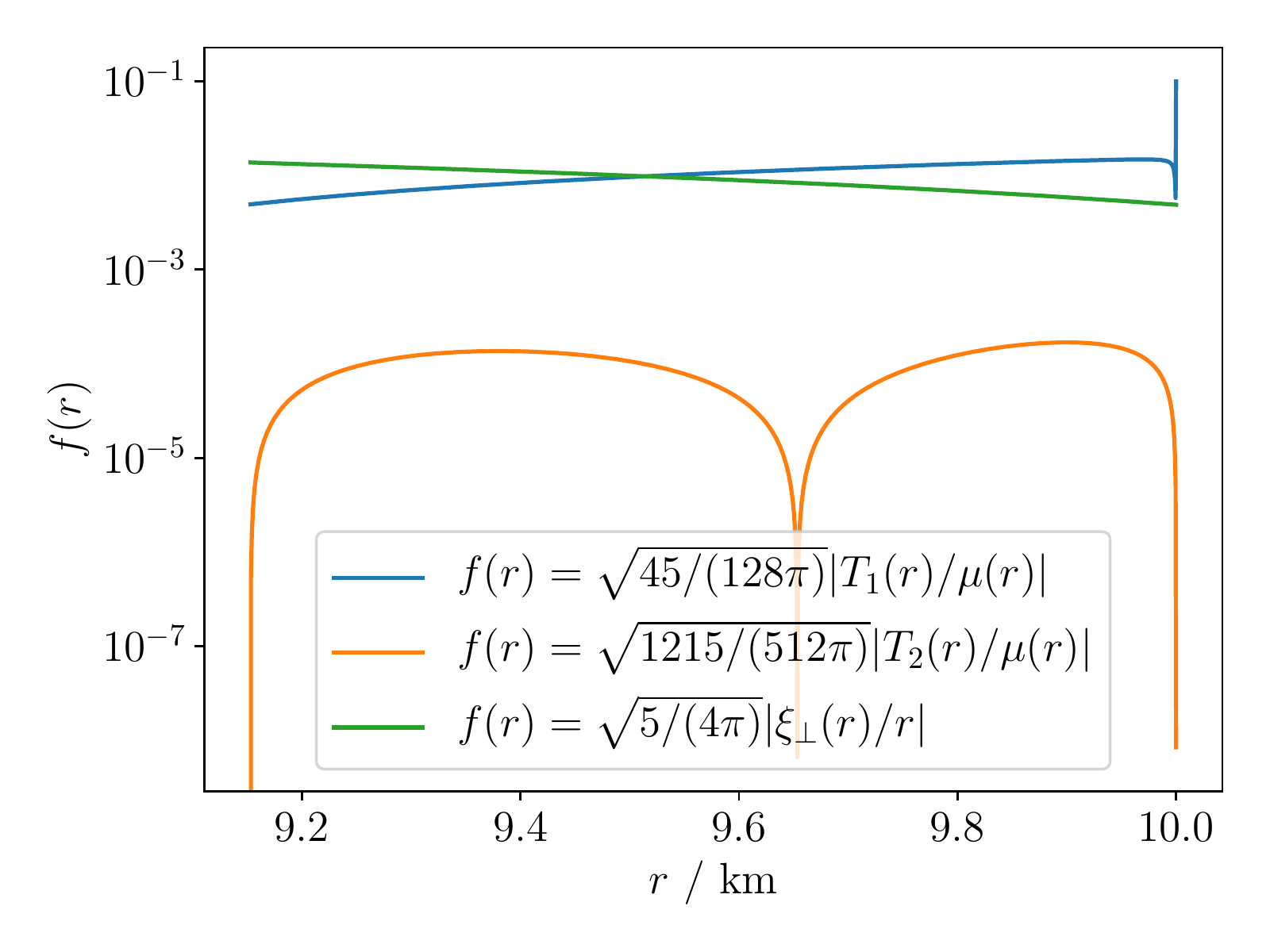}%
    \caption{\label{fig:Strain}The strain components in (\ref{eq:vonMises}) 
             maximised over $(\theta, \phi)$ against radius for the potential 
             solution to Laplace's equation.}
\end{figure}

Thus, for the strained star (star C) we integrate 
equations~(\ref{eqs:FluidPerturbations}) for 
the core and ocean and integrate equations~(\ref{eqs:ElasticPerturbations}) in 
the 
elastic crust. The relaxed star (star A) is generated using 
equations~(\ref{eqs:FluidPerturbations}) for the entire star. The perturbations 
are 
normalised by ensuring that the point in the crust where the strain is highest 
reaches breaking strain, according to (\ref{eq:vonMises}). 
The force associated with this deformation (\ref{eq:SourcePotentialSurface}) is 
then taken to be the same for the relaxed star. Figs.~\ref{fig:Traction} and 
\ref{fig:Strain} show the results for the strained star. In 
Fig.~\ref{fig:Traction} we show how the perturbed traction is continuous at the 
fluid-elastic interfaces. We note that Fig.~\ref{fig:Strain} shows how the 
dominant contribution to the von Mises strain comes from the radial traction 
component. This is also true for the other forces we consider. It is at the top 
of the crust that the star is the weakest in the $(l, m) = (2, 2)$ mode. The 
quadrupoles are calculated using (\ref{eq:Multipole3}). The relaxed star attains 
a quadrupole of
$|Q_{2 2}^\text{relaxed}| = \SI{2.4e43}{\gram\centi\metre\squared}$, which 
corresponds to an ellipticity of 
$|\epsilon^\text{relaxed}| = \num{3.1e-2}$. The difference between the strained 
and relaxed star is $|Q_{2 2}^\text{strained} - Q_{2 2}^\text{relaxed}| 
= \SI{1.7e37}{\gram\centi\metre\squared}$, 
$|\epsilon^\text{strained} - \epsilon^\text{relaxed}| = \num{2.2e-8}$. 

The very different sizes of $|\epsilon^\text{relaxed}|$ and 
$|\epsilon^\text{strained} - \epsilon^\text{relaxed}|$ reported in 
Table~\ref{tab:ellipticities} have a natural interpretation. The large 
ellipticity represented by $|\epsilon^\text{relaxed}|$ corresponds to a star 
whose deformation is supported by the external force $f_i$, with a size limited 
only by the crustal breaking strain. In this case, the (non-zero) shear modulus 
of the crust plays little role. [It is this sort of configuration that was 
effectively considered in \citet{2006MNRAS.373.1423H}, where in that case the 
force that was implicitly introduced was a force per unit area, applied at the 
surface.] In contrast, the ellipticity represented by 
$|\epsilon^\text{strained} - \epsilon^\text{relaxed}|$ is that supported by the 
shear strains of the crust when the external force is removed, and therefore is 
sensitive to the crust's shear modulus. As is readily captured by simple 
back-of-the-envelope estimates, the relative sizes of these two ellipticities 
are related to the fact that the gravitational binding energy of the star is 
orders of magnitude larger than the Coulomb binding energy of the crustal 
lattice [see, e.g., \citet{2002CQGra..19.1255J}].

We observe that the ellipticity
$|\epsilon^\text{strained} - \epsilon^\text{relaxed}| = \num{2.2e-8}$ is notably 
smaller than what has been found in previous work 
[equations~(\ref{eq:Q_22UCB})--(\ref{eq:Q_22JMO})]. This is not surprising, as 
these 
previous studies considered strain fields that were maximal everywhere, as 
opposed to at a single point. With a view to producing larger ellipticities, we 
will therefore consider some different choices of external force field.

\subsection{A solution of Laplace's equation outside the core}

We consider a special case of the above source: a source potential that does not 
act in the core --  instead, it only manifests itself in the crust and ocean. The 
motivation for considering this special case is regularity at the centre will 
not be a necessary condition on the source potential since it does not exist at 
the origin. Also, we note that \citet{2006MNRAS.373.1423H} found that a similar 
example produced their largest quadrupole moment. We then have the general 
solution to Laplace's equation (\ref{eq:Laplaces}), 
\begin{equation}
    \chi(r) = A r^l + B / r^{l + 1}, 
\label{eq:SourcePotential}
\end{equation}
where $B$ is another constant. This expression is taken to be true for the base 
of the crust and above. 

As this  model is somewhat artificial, we have to make a number of 
assumptions with regards to its prescription. We take the core to be 
unperturbed and have $\delta \Phi = \xi_r = 0$ in the core. With the 
introduction of the source potential in the crust, there will be a discontinuity 
in $U$ at the core-crust interface. However, we insist that $\delta \Phi$ must 
be continuous. This discontinuity is relevant for the radial traction 
condition~(\ref{eq:RadialTraction}) where $U_\text{F} = 0$, but has a finite 
value in the crust due to the source potential. 

The quadrupole may be calculated from (\ref{eq:Multipole2}). The matching with 
the total perturbed potential needs to be adjusted to take into account the 
additional $1 / r^{l + 1}$ term from the external field. Therefore, we have 
\begin{equation}
    Q_{l m} = \frac{R^{l + 1}}{4 \uppi G} [R U'(R) - l U(R)] 
        + \frac{2 l + 1}{4 \uppi G} B. 
\end{equation}

\begin{figure*}
	\includegraphics[width=0.49\textwidth]{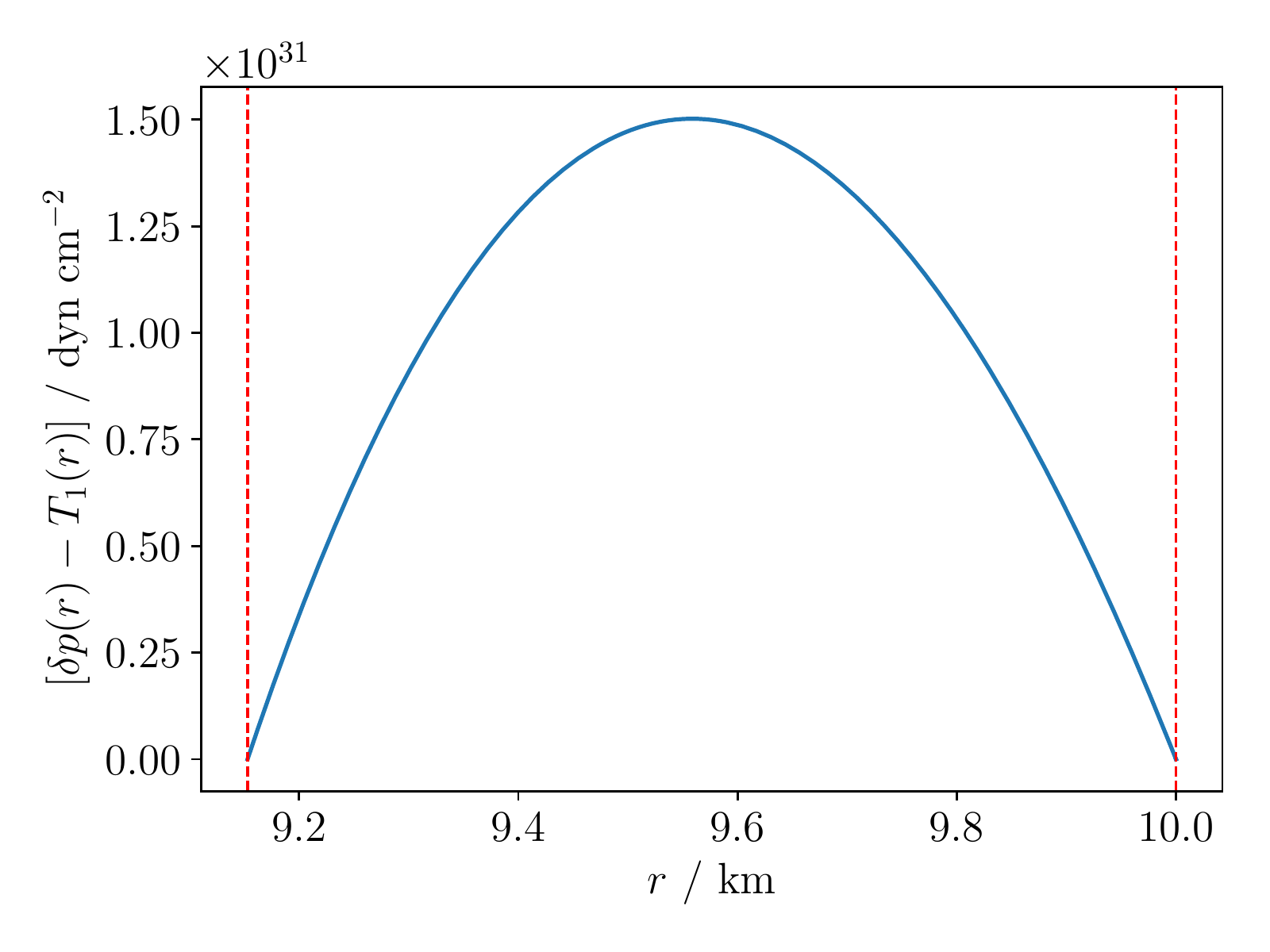}%
    \includegraphics[width=0.49\textwidth]{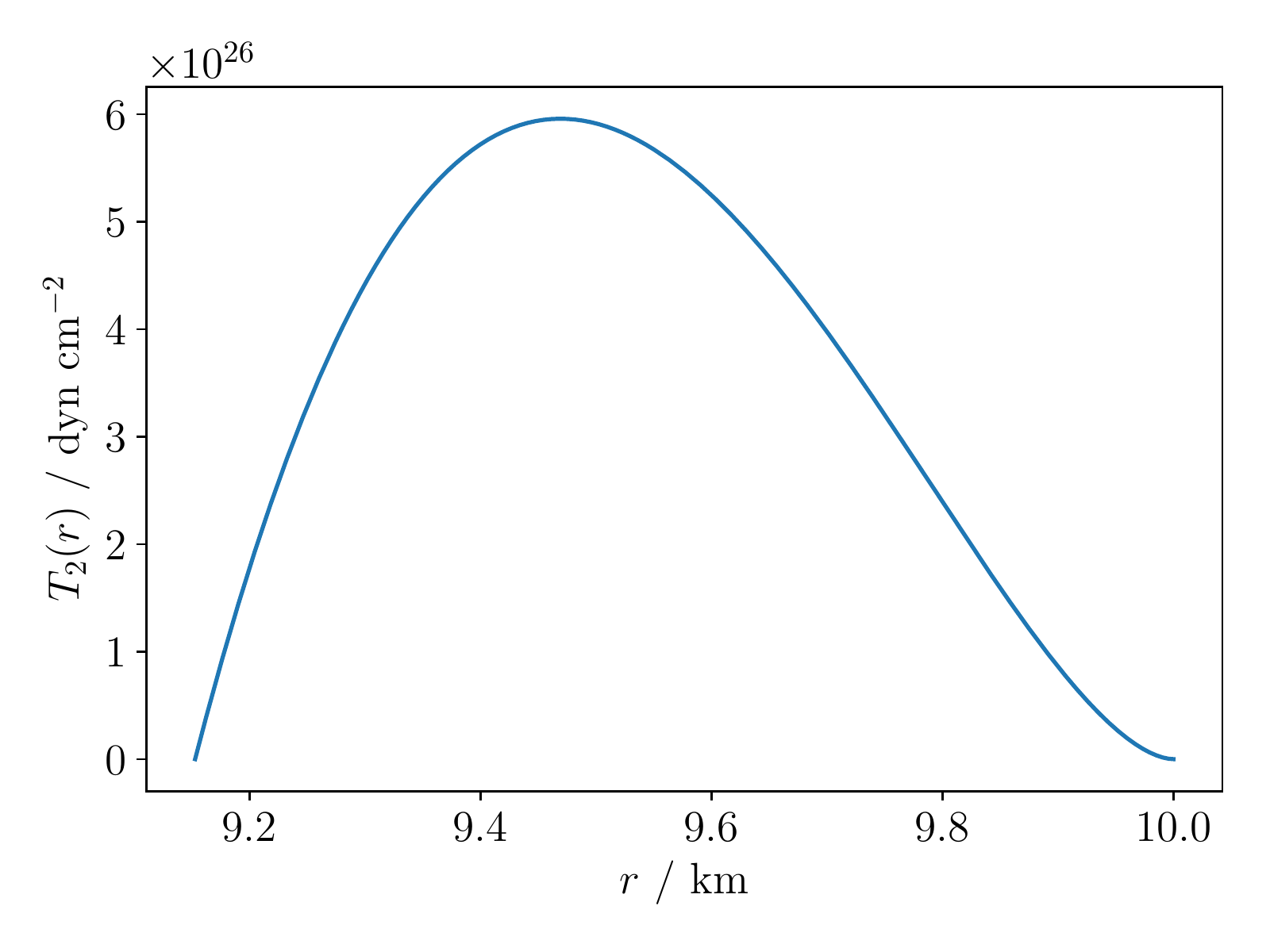}%
    \caption{\label{fig:TractionSpecial}The radial (left panel) and tangential 
             (right panel) components of the perturbed traction as functions of 
             radius for the potential solution to Laplace's equation outside the 
             core. The vertical red dashed lines in the left panel indicate the 
             base and the top of the crust. Regarding the horizontal range in 
             the right panel, recall that $T_2$ only has a finite value in the 
             crust.}
\end{figure*}

\begin{figure}
	\includegraphics[width=\columnwidth]{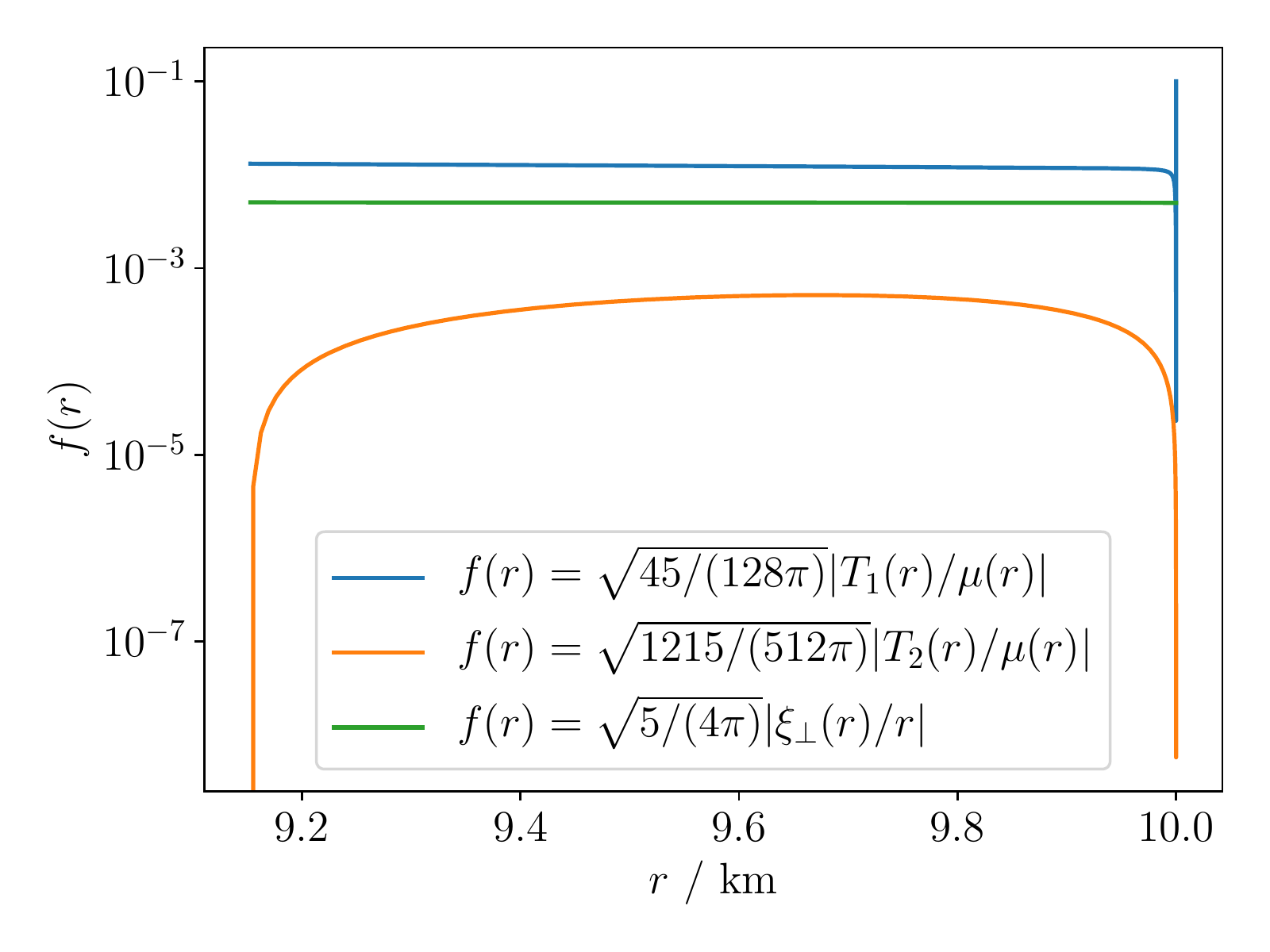}%
    \caption{\label{fig:StrainSpecial}The strain components in 
             (\ref{eq:vonMises}) maximised over $(\theta, \phi)$ against 
             radius for the potential solution to Laplace's equation outside the 
             core.}
\end{figure}

As in the previous case, we generate a relaxed star and a maximally strained 
star. One must vary either $A$ or $B$ to ensure the surface boundary 
condition~(\ref{eq:PerturbedPotentialSurface}) is satisfied. We normalise the 
relaxed star so that it experiences the same source 
potential~(\ref{eq:SourcePotential}). The 
results for this case are shown in Figs.~\ref{fig:TractionSpecial} and 
\ref{fig:StrainSpecial}. As in the above example, the $T_1$ component dominates 
the von Mises strain and the crust breaks at the top. We find 
$|Q_{2 2}^\text{relaxed}| = \SI{1.4e41}{\gram\centi\metre\squared}$, 
$|\epsilon^\text{relaxed}| = \num{1.8e-4}$ and 
$|Q_{2 2}^\text{strained} - Q_{2 2}^\text{relaxed}| 
= \SI{4.4e38}{\gram\centi\metre\squared}$, 
$|\epsilon^\text{strained} - \epsilon^\text{relaxed}| = \num{5.7e-7}$. 

Compared to the previous result, the quadrupole difference between the 
relaxed and strained stars  has increased by an order of magnitude. This is 
within a factor of a few of previous maximum-mountain calculations, and 
illustrates the dependence on the force prescription.

\subsection{A thermal pressure perturbation}
\label{sec:ThermalPressure}

\begin{figure*}
	\includegraphics[width=0.49\textwidth]{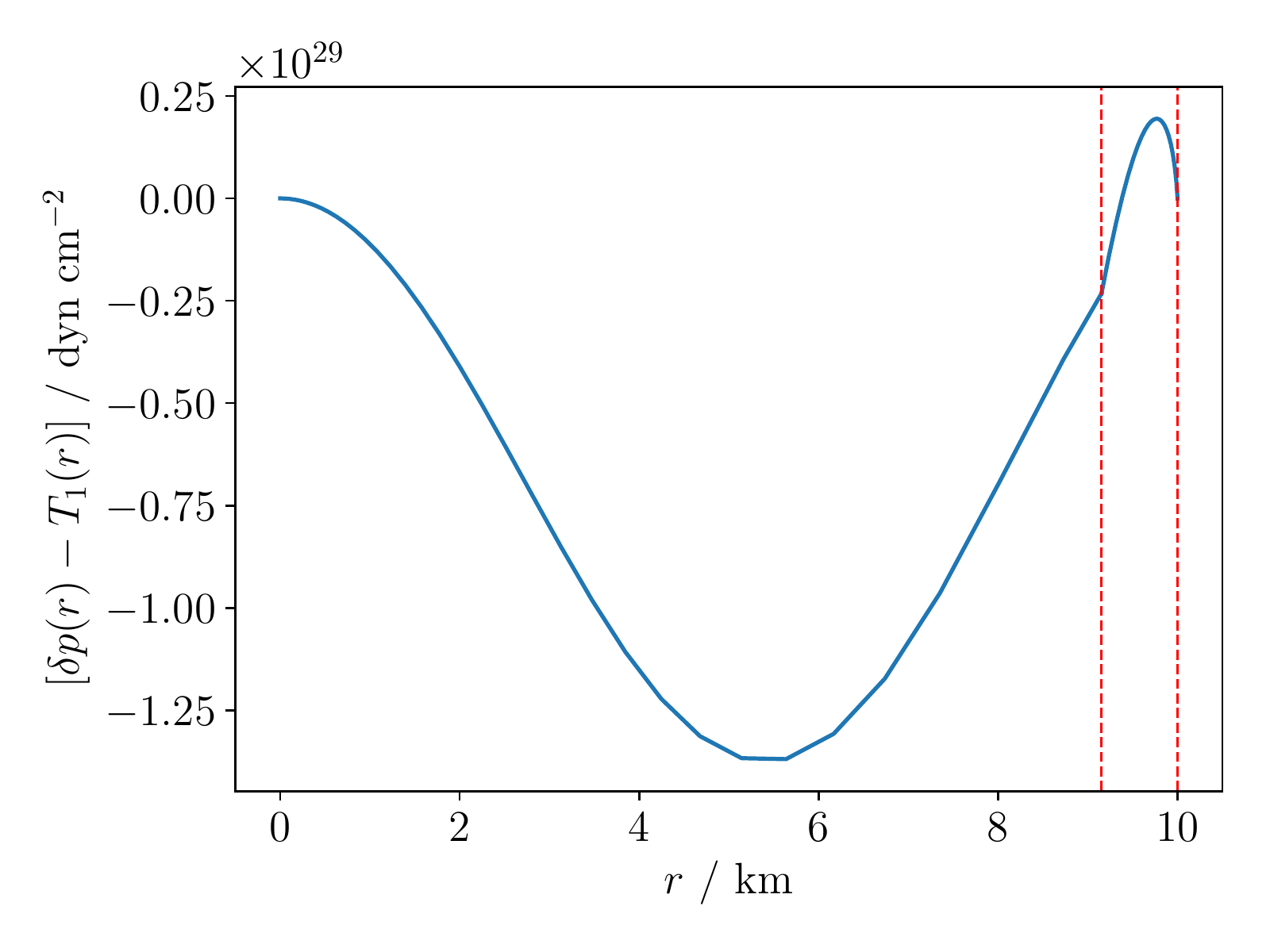}%
    \includegraphics[width=0.49\textwidth]{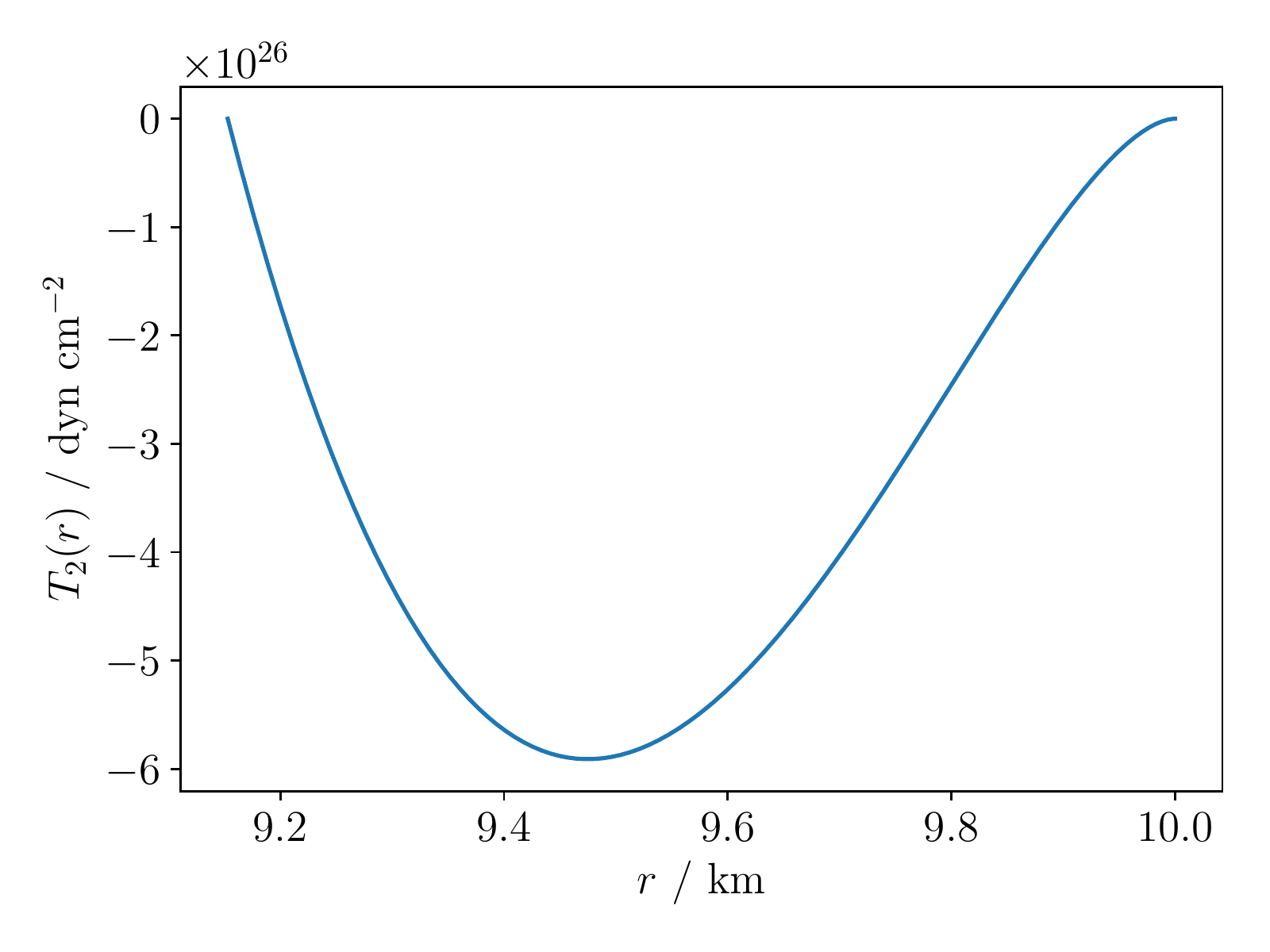}%
    \caption{\label{fig:TractionTemperature}The radial (left panel) and 
             tangential (right panel) components of the perturbed traction as 
             functions of radius for the temperature perturbation. The vertical 
             red dashed lines in the left panel indicate the base and the top of 
             the crust. Regarding the horizontal range in the right panel, 
             recall that $T_2$ only has a finite value in the crust.}
\end{figure*}

\begin{figure}
	\includegraphics[width=\columnwidth]{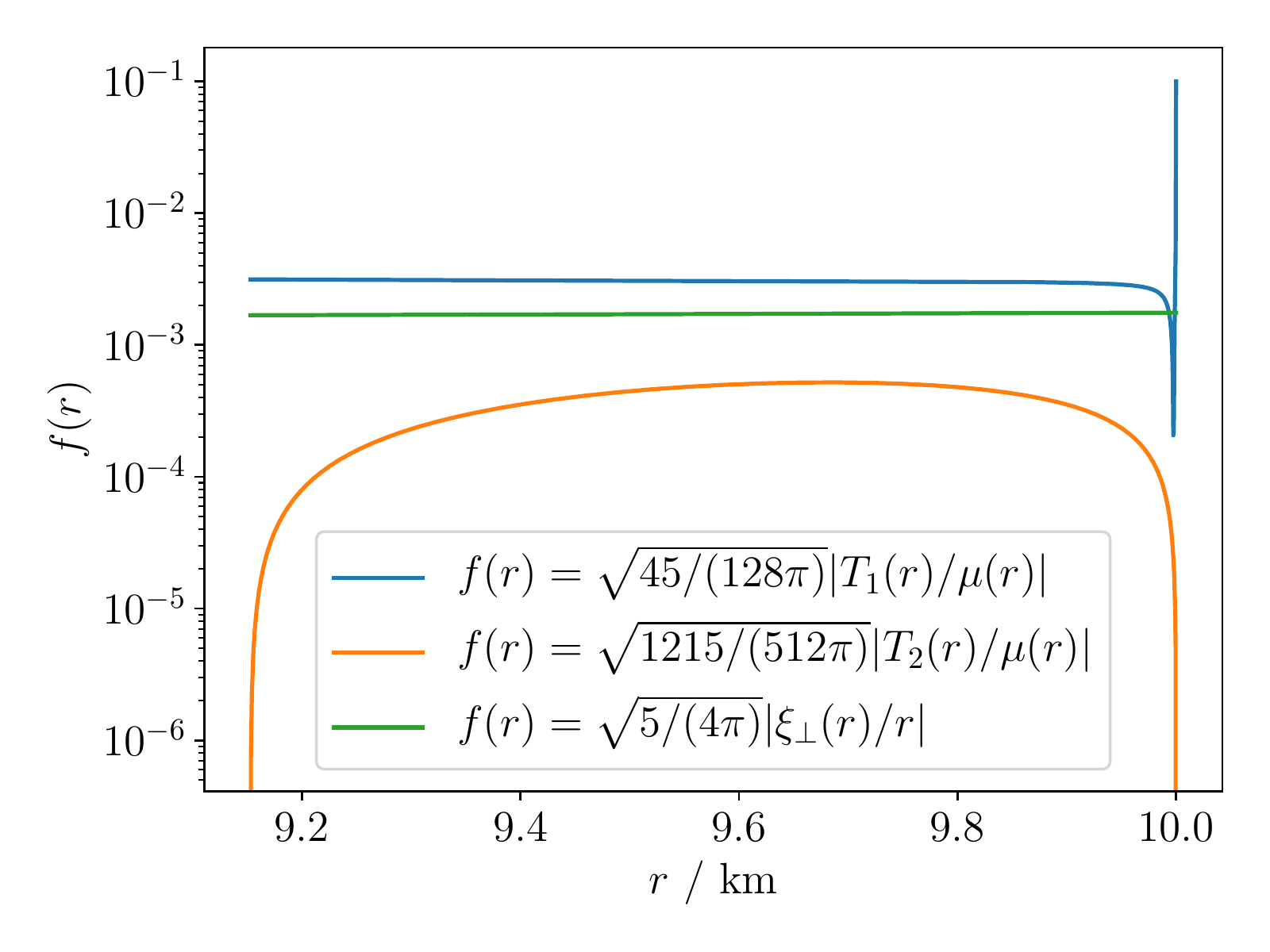}%
    \caption{\label{fig:StrainTemperature}The strain components in 
             (\ref{eq:vonMises}) maximised over $(\theta, \phi)$ against 
             radius for the temperature perturbation.}
\end{figure}

The third source for the perturbations we examine is motivated by a thermal 
pressure perturbation. Note that the approach we use for this example 
could be applied more generally to consider non-barotropic matter where the 
pressure is adjusted, relative to the barotropic case, at the perturbative 
level. 
We assume the thermal pressure to be of the ideal-gas form, 
\begin{equation}
    \delta p_\text{th} = \frac{k_\text{B} \rho}{m_\text{u}} \delta T, 
\end{equation}
where $k_\text{B}$ is the Boltzmann constant, $m_\text{u}$ is the atomic mass 
unit and $\delta T$ is the temperature perturbation. To interpret this thermal 
pressure as a force, we identify 
\begin{equation}
    \rho \nabla_i \chi = \nabla_i \delta p_\text{th} 
        = \frac{k_\text{B}}{m_\text{u}} \nabla_i (\rho \delta T). 
\end{equation}
The temperature perturbation must be regular at the origin. For simplicity we 
assume it to be quadratic, 
\begin{equation}
    \delta T(r) = \left( \frac{r}{R} \right)^2 \delta T(R), 
\end{equation}
where $\delta T(R)$ corresponds to the perturbation of the temperature at the 
surface. Both the relaxed and strained configurations experience the same 
temperature perturbation. We show the results in 
Figs.~\ref{fig:TractionTemperature} and \ref{fig:StrainTemperature}. We now find 
the crust breaks when $\delta T(R) = \SI{3.5e6}{\kelvin}$. We obtain the  
results $|Q_{2 2}^\text{relaxed}| = \SI{9.2e38}{\gram\centi\metre\squared}$, 
$|\epsilon^\text{relaxed}| = \num{1.2e-6}$ and 
$|Q_{2 2}^\text{strained} - Q_{2 2}^\text{relaxed}| 
= \SI{4.0e38}{\gram\centi\metre\squared}$, 
$|\epsilon^\text{strained} - \epsilon^\text{relaxed}| = \num{5.2e-7}$. This 
result is of the same order of magnitude to the potential outside the core.

It is interesting to note that while the values of the ellipticities 
$|\epsilon^\text{strained}|$ and $|\epsilon^\text{relaxed}|$ vary by about four 
orders of magnitude for the three deforming forces we consider, the variation in 
the actual ellipticity of the mountain, 
$|\epsilon^\text{strained} - \epsilon^\text{relaxed}|$, is relatively modest, 
about one order of magnitude (see Table~\ref{tab:ellipticities}). This is 
presumably a reflection of the fact that in all three cases we consider the same 
star with the same crustal breaking strain and shear modulus, so all stars have 
a similar ability to support deformations.

\section{Conclusions}
\label{sec:Conclusions}

The question of the maximum mountain a neutron star crust can support is an 
interesting problem. Such an estimate provides upper limits on the strength of 
gravitational-wave emission from rotating neutron stars, as well as having 
implications for the maximum spin-frequency limit that these systems can attain. 

We returned to this problem to tackle some of the pertinent assumptions made in 
previous work. We have discussed how previous estimates have not dealt 
appropriately with boundary conditions that must be satisfied for realistic 
neutron star models. The calculations of \citet{2000MNRAS.319..902U} and 
\citet{2013PhRvD..88d4004J} both assumed a specific form for the strain that 
takes the star away from its relaxed shape and ensures the crust is maximally 
strained at every point. However, such a strain is somewhat unphysical since it 
does not respect the continuity of the traction vector.  Additionally, the 
approach of \citet{2006MNRAS.373.1423H}, while satisfying the traction 
conditions at the crust-core boundary, did not obey the boundary condition on 
the potential at the surface. This was due to the calculation assuming the 
relaxed configuration is spherical and implicitly using a surface force to 
deform the star. There were also errors present in the perturbation equations of 
\citet{2006MNRAS.373.1423H} which change their results by several orders of 
magnitude. 

An important simplification of the previous studies was to not explicitly 
calculate the non-spherical, relaxed shape that the strain is taken with respect 
to. As we have shown, such a description requires the introduction of a 
perturbing force which takes the star away from sphericity. We found such a 
discussion was missing in prior studies and, hence, have provided a 
demonstration that shows, provided one has a description of the strain, how the 
relaxed shape can calculated.

We found that including this force is crucial in enabling one to satisfy all the 
boundary conditions. Therefore, we have introduced a novel scheme for 
calculating the maximum quadrupole deformation that a neutron star can sustain 
and have demonstrated how our scheme is entirely equivalent to the approach of 
preceding calculations. Crucially, the formalism satisfies all the boundary 
conditions of the problem. One of the key advantages of our approach is that one 
computes all relevant quantities, including the shape of the relaxed star. 
However, one must provide a prescription for the deforming force. 

There is obviously significant freedom in what one may choose for the form of 
this force and, indeed, the formalism we have presented can be used for 
any deforming force that has the form (\ref{eq:Force}). Furthermore, 
it would not be difficult to adjust this formalism for other forces. 
However, evolutionary calculations will be necessary to fully motivate 
the form of the force. Thus, we surveyed three simple examples for the 
source of the mountains. We obtained 
the largest quadrupole for the (somewhat artificial) case where the perturbing 
potential is a solution to Laplace's equation, but leaves the core unperturbed. 
All of our results are between a factor of a few to two orders of magnitude 
below that of prior estimates for the maximum mountain a neutron star may 
support. That our results were smaller is not surprising, as our maximum 
mountains were constructed so that the breaking strain was reached at only a 
single point. An immediate question would be if there is a reasonable scenario 
that bridges the gap between relaxed configurations associated with a specific 
force and configurations following from specifying the strain. It seems 
inevitable that the answer will rely on evolutionary scenarios, leading to 
mountain formation, a problem that has not yet attracted the attention it 
deserves. 

An example of a promising scenario through which a rotating neutron star 
may radiate gravitational waves is accretion from a binary companion. As the gas 
is accreted onto the surface of the star, chemical reactions take place which 
change the composition. Such changes in the composition can, in turn, result in 
the star attaining a non-trivial quadrupole moment 
\citep{1998ApJ...501L..89B, 2000MNRAS.319..902U}. Additionally, there has been 
some effort towards calculating mountains on accreting neutron stars that are 
sustained by the magnetic field 
\citep{2005ApJ...623.1044M, 2006ApJ...641..471P, 2011MNRAS.417.2696P}. 
We note that, in our calculation, we only consider barotropic matter. This is 
appropriate to describe equilibrium stellar models. Indeed, if the star is in 
equilibrium, accreted and non-accreted matter may be described using barotropic 
equations of state \citep{2006MNRAS.373.1423H}. However, for evolutionary 
calculations, like those described above, one may need to consider 
non-barotropic features and, as we noted in 
Section~\ref{sec:ThermalPressure}, the formalism we have presented could be used 
with such aspects at the perturbative level.

As an (admittedly phenomenological) indication of a possible solution, it may be 
worth pointing out that our approach to elasticity is somewhat simplistic. We 
have followed the usual assumption that the crust can be well described as an 
elastic solid (represented by a linear stress-strain relation) until it 
reaches the breaking strain, at which point the crust fails and all the strain 
is released. This model accords well with the molecular dynamics simulations of 
\citet{2009PhRvL.102s1102H}, but it is worth noting that laboratory materials 
tends to behave slightly differently \citep{75da23d2ccdd4ab1b1b32564cc0df76b}. 
In 
particular, one typically finds that material deforms plastically for some level 
of strain before the ultimate failure. This introduces the yield strain as the 
point above which the stress-strain relationship is no longer linear and raises 
(difficult) questions regarding the plastic behaviour (the matter may 
harden, allowing stresses to continue building, or soften, leading to reduced 
stress as the strain increases). State-of-the-art simulations suggest a narrow 
region of plastic behaviour before the crust fails \citep{2009PhRvL.102s1102H}, 
but one should perhaps keep in mind that the levels of shear involved in the 
simulation may not lead to a true representation of matter that is deformed more 
gently. Let us, for the sake of the argument, suppose that this is the case and 
that the crust exhibits ideal plasticity above then chosen yield strain. If this 
were to happen, the strain would locally saturate at the yield limit even if the 
imposed force increased. One can then imagine applying a deforming force to 
source a neutron star mountain and then increasing it until some point in the 
crust reaches yield strain. This is essentially the calculation we have done, as 
we did not model the behaviour beyond this point. Allowing for (ideal) plastic 
flow as the force is further increased, one may envisage that the entire crust 
may saturate at the yield strain. This is, of course, pure speculation [although 
there have been several notable discussions about the relevance of plastic 
deformations of the neutron star crust; see 
\citet{1970PhRvL..24.1191S}, \citet{2003ApJ...595..342J} and 
\citet{2010MNRAS.407L..54C}], but it 
might explain how a real system could reach the maximum strain configuration 
imposed in the \citet{2000MNRAS.319..902U} argument. As we already suggested, 
detailed evolutionary calculations which take into account the physical 
processes that produce the mountain will be required to make progress on the 
problem. 

Another natural avenue for future research is to generalise our calculation to 
relativity. This should be reasonably straightforward to do. One would need to 
use the relativistic equivalents of the perturbation equations [see, 
e.g., \citet{2020PhRvD.101j3025G}]. This would be an important step as 
it brings realistic equations of state into play. 

\section*{Acknowledgements}

N.A. and D.I.J. are grateful for financial support from STFC via Grant No. 
ST/R00045X/1. 

\section*{Data availability}

Additional data underlying this article will be shared on reasonable request to 
the corresponding author. 


\bibliographystyle{mnras}
\bibliography{bibliography}

\providecommand{\noopsort}[1]{}\providecommand{\singleletter}[1]{#1}%
\begin{thebibliography}{}
\makeatletter
\relax
\def\mn@urlcharsother{\let\do\@makeother \do\$\do\&\do\#\do\^\do\_\do\%\do\~}
\def\mn@doi{\begingroup\mn@urlcharsother \@ifnextchar [ {\mn@doi@}
  {\mn@doi@[]}}
\def\mn@doi@[#1]#2{\def\@tempa{#1}\ifx\@tempa\@empty \href
  {http://dx.doi.org/#2} {doi:#2}\else \href {http://dx.doi.org/#2} {#1}\fi
  \endgroup}
\def\mn@eprint#1#2{\mn@eprint@#1:#2::\@nil}
\def\mn@eprint@arXiv#1{\href {http://arxiv.org/abs/#1} {{\tt arXiv:#1}}}
\def\mn@eprint@dblp#1{\href {http://dblp.uni-trier.de/rec/bibtex/#1.xml}
  {dblp:#1}}
\def\mn@eprint@#1:#2:#3:#4\@nil{\def\@tempa {#1}\def\@tempb {#2}\def\@tempc
  {#3}\ifx \@tempc \@empty \let \@tempc \@tempb \let \@tempb \@tempa \fi \ifx
  \@tempb \@empty \def\@tempb {arXiv}\fi \@ifundefined
  {mn@eprint@\@tempb}{\@tempb:\@tempc}{\expandafter \expandafter \csname
  mn@eprint@\@tempb\endcsname \expandafter{\@tempc}}}

\bibitem[\protect\citeauthoryear{{Aasi} et~al.}{{Aasi}
  et~al.}{2013}]{2013PhRvD..87d2001A}
{Aasi} J.,  et~al., 2013, \mn@doi [\prd] {10.1103/PhysRevD.87.042001}, \href
  {http://adsabs.harvard.edu/abs/2013PhRvD..87d2001A} {87, 042001}

\bibitem[\protect\citeauthoryear{{Aasi} et~al.}{{Aasi}
  et~al.}{2014}]{2014ApJ...785..119A}
{Aasi} J.,  et~al., 2014, \mn@doi [\apj] {10.1088/0004-637X/785/2/119}, \href
  {http://adsabs.harvard.edu/abs/2014ApJ...785..119A} {785, 119}

\bibitem[\protect\citeauthoryear{{Aasi} et~al.}{{Aasi}
  et~al.}{2015a}]{2015PhRvD..91b2004A}
{Aasi} J.,  et~al., 2015a, \mn@doi [\prd] {10.1103/PhysRevD.91.022004}, \href
  {http://adsabs.harvard.edu/abs/2015PhRvD..91b2004A} {91, 022004}

\bibitem[\protect\citeauthoryear{{Aasi} et~al.}{{Aasi}
  et~al.}{2015b}]{2015PhRvD..91f2008A}
{Aasi} J.,  et~al., 2015b, \mn@doi [\prd] {10.1103/PhysRevD.91.062008}, \href
  {http://adsabs.harvard.edu/abs/2015PhRvD..91f2008A} {91, 062008}

\bibitem[\protect\citeauthoryear{{Abadie} et~al.}{{Abadie}
  et~al.}{2010}]{2010CQGra..27q3001A}
{Abadie} J.,  et~al., 2010, \mn@doi [Class. Quantum Gravity]
  {10.1088/0264-9381/27/17/173001}, \href
  {https://ui.adsabs.harvard.edu/abs/2010CQGra..27q3001A} {27, 173001}

\bibitem[\protect\citeauthoryear{{Abadie} et~al.}{{Abadie}
  et~al.}{2011a}]{2011PhRvD..83d2001A}
{Abadie} J.,  et~al., 2011a, \mn@doi [\prd] {10.1103/PhysRevD.83.042001}, \href
  {http://adsabs.harvard.edu/abs/2011PhRvD..83d2001A} {83, 042001}

\bibitem[\protect\citeauthoryear{{Abadie} et~al.}{{Abadie}
  et~al.}{2011b}]{2011ApJ...737...93A}
{Abadie} J.,  et~al., 2011b, \mn@doi [\apj] {10.1088/0004-637X/737/2/93}, \href
  {http://adsabs.harvard.edu/abs/2011ApJ...737...93A} {737, 93}

\bibitem[\protect\citeauthoryear{{Abadie} et~al.}{{Abadie}
  et~al.}{2012}]{2012PhRvD..85b2001A}
{Abadie} J.,  et~al., 2012, \mn@doi [\prd] {10.1103/PhysRevD.85.022001}, \href
  {http://adsabs.harvard.edu/abs/2012PhRvD..85b2001A} {85, 022001}

\bibitem[\protect\citeauthoryear{{Abbott} et~al.}{{Abbott}
  et~al.}{2004}]{2004PhRvD..69h2004A}
{Abbott} B.,  et~al., 2004, \mn@doi [\prd] {10.1103/PhysRevD.69.082004}, \href
  {http://adsabs.harvard.edu/abs/2004PhRvD..69h2004A} {69, 082004}

\bibitem[\protect\citeauthoryear{{Abbott} et~al.}{{Abbott}
  et~al.}{2005a}]{2005PhRvD..72j2004A}
{Abbott} B.,  et~al., 2005a, \mn@doi [\prd] {10.1103/PhysRevD.72.102004}, \href
  {http://adsabs.harvard.edu/abs/2005PhRvD..72j2004A} {72, 102004}

\bibitem[\protect\citeauthoryear{{Abbott} et~al.}{{Abbott}
  et~al.}{2005b}]{2005PhRvL..94r1103A}
{Abbott} B.,  et~al., 2005b, \mn@doi [\prl] {10.1103/PhysRevLett.94.181103},
  \href {http://adsabs.harvard.edu/abs/2005PhRvL..94r1103A} {94, 181103}

\bibitem[\protect\citeauthoryear{{Abbott} et~al.}{{Abbott}
  et~al.}{2007a}]{2007PhRvD..76d2001A}
{Abbott} B.,  et~al., 2007a, \mn@doi [\prd] {10.1103/PhysRevD.76.042001}, \href
  {http://adsabs.harvard.edu/abs/2007PhRvD..76d2001A} {76, 042001}

\bibitem[\protect\citeauthoryear{{Abbott} et~al.}{{Abbott}
  et~al.}{2007b}]{2007PhRvD..76h2001A}
{Abbott} B.,  et~al., 2007b, \mn@doi [\prd] {10.1103/PhysRevD.76.082001}, \href
  {http://adsabs.harvard.edu/abs/2007PhRvD..76h2001A} {76, 082001}

\bibitem[\protect\citeauthoryear{{Abbott} et~al.}{{Abbott}
  et~al.}{2008a}]{2008PhRvD..77b2001A}
{Abbott} B.,  et~al., 2008a, \mn@doi [\prd] {10.1103/PhysRevD.77.022001}, \href
  {http://adsabs.harvard.edu/abs/2008PhRvD..77b2001A} {77, 022001}

\bibitem[\protect\citeauthoryear{{Abbott} et~al.}{{Abbott}
  et~al.}{2008b}]{2008ApJ...683L..45A}
{Abbott} B.,  et~al., 2008b, \mn@doi [\apjl] {10.1086/591526}, \href
  {http://adsabs.harvard.edu/abs/2008ApJ...683L..45A} {683, L45}

\bibitem[\protect\citeauthoryear{{Abbott} et~al.}{{Abbott}
  et~al.}{2009}]{2009PhRvD..79b2001A}
{Abbott} B.,  et~al., 2009, \mn@doi [\prd] {10.1103/PhysRevD.79.022001}, \href
  {http://adsabs.harvard.edu/abs/2009PhRvD..79b2001A} {79, 022001}

\bibitem[\protect\citeauthoryear{{Abbott} et~al.}{{Abbott}
  et~al.}{2010}]{2010ApJ...713..671A}
{Abbott} B.~P.,  et~al., 2010, \mn@doi [\apj] {10.1088/0004-637X/713/1/671},
  \href {http://adsabs.harvard.edu/abs/2010ApJ...713..671A} {713, 671}

\bibitem[\protect\citeauthoryear{{Abbott} et~al.}{{Abbott}
  et~al.}{2016}]{2016PhRvD..94d2002A}
{Abbott} B.~P.,  et~al., 2016, \mn@doi [\prd] {10.1103/PhysRevD.94.042002},
  \href {http://adsabs.harvard.edu/abs/2016PhRvD..94d2002A} {94, 042002}

\bibitem[\protect\citeauthoryear{{Abbott} et~al.}{{Abbott}
  et~al.}{2017a}]{2017PhRvD..95l2003A}
{Abbott} B.~P.,  et~al., 2017a, \mn@doi [\prd] {10.1103/PhysRevD.95.122003},
  \href {http://adsabs.harvard.edu/abs/2017PhRvD..95l2003A} {95, 122003}

\bibitem[\protect\citeauthoryear{{Abbott} et~al.}{{Abbott}
  et~al.}{2017b}]{2017PhRvD..96f2002A}
{Abbott} B.~P.,  et~al., 2017b, \mn@doi [\prd] {10.1103/PhysRevD.96.062002},
  \href {http://adsabs.harvard.edu/abs/2017PhRvD..96f2002A} {96, 062002}

\bibitem[\protect\citeauthoryear{{Abbott} et~al.}{{Abbott}
  et~al.}{2017c}]{2017PhRvD..96l2006A}
{Abbott} B.~P.,  et~al., 2017c, \mn@doi [\prd] {10.1103/PhysRevD.96.122006},
  \href {http://adsabs.harvard.edu/abs/2017PhRvD..96l2006A} {96, 122006}

\bibitem[\protect\citeauthoryear{Abbott et~al.}{Abbott
  et~al.}{2017d}]{2017PhRvL.119p1101A}
Abbott B.~P.,  et~al., 2017d, \mn@doi [\prl] {10.1103/PhysRevLett.119.161101},
  \href {https://ui.adsabs.harvard.edu/abs/2017PhRvL.119p1101A} {119, 161101}

\bibitem[\protect\citeauthoryear{{Abbott} et~al.}{{Abbott}
  et~al.}{2017e}]{2017ApJ...839...12A}
{Abbott} B.~P.,  et~al., 2017e, \mn@doi [\apj] {10.3847/1538-4357/aa677f},
  \href {http://adsabs.harvard.edu/abs/2017ApJ...839...12A} {839, 12}

\bibitem[\protect\citeauthoryear{{Abbott} et~al.}{{Abbott}
  et~al.}{2017f}]{2017ApJ...847...47A}
{Abbott} B.~P.,  et~al., 2017f, \mn@doi [\apj] {10.3847/1538-4357/aa86f0},
  \href {http://adsabs.harvard.edu/abs/2017ApJ...847...47A} {847, 47}

\bibitem[\protect\citeauthoryear{{Abbott} et~al.}{{Abbott}
  et~al.}{2018a}]{2018PhRvD..97j2003A}
{Abbott} B.~P.,  et~al., 2018a, \mn@doi [\prd] {10.1103/PhysRevD.97.102003},
  \href {https://ui.adsabs.harvard.edu/abs/2018PhRvD..97j2003A} {97, 102003}

\bibitem[\protect\citeauthoryear{{Abbott} et~al.}{{Abbott}
  et~al.}{2018b}]{2018PhRvL.120c1104A}
{Abbott} B.~P.,  et~al., 2018b, \mn@doi [\prl]
  {10.1103/PhysRevLett.120.031104}, \href
  {http://adsabs.harvard.edu/abs/2018PhRvL.120c1104A} {120, 031104}

\bibitem[\protect\citeauthoryear{{Abbott} et~al.}{{Abbott}
  et~al.}{2019a}]{2019PhRvD..99l2002A}
{Abbott} B.~P.,  et~al., 2019a, \mn@doi [\prd] {10.1103/PhysRevD.99.122002},
  \href {https://ui.adsabs.harvard.edu/abs/2019PhRvD..99l2002A} {99, 122002}

\bibitem[\protect\citeauthoryear{Abbott et~al.}{Abbott
  et~al.}{2019b}]{PhysRevD.100.122002}
Abbott B.~P.,  et~al., 2019b, \mn@doi [\prd] {10.1103/PhysRevD.100.122002},
  \href {https://ui.adsabs.harvard.edu/abs/2019arXiv190612040T} {100, 122002}

\bibitem[\protect\citeauthoryear{{Abbott} et~al.}{{Abbott}
  et~al.}{2019c}]{2019PhRvD.100b4004A}
{Abbott} B.~P.,  et~al., 2019c, \mn@doi [\prd] {10.1103/PhysRevD.100.024004},
  \href {https://ui.adsabs.harvard.edu/abs/2019PhRvD.100b4004A} {100, 024004}

\bibitem[\protect\citeauthoryear{{Abbott} et~al.}{{Abbott}
  et~al.}{2019d}]{2019ApJ...875..122A}
{Abbott} B.~P.,  et~al., 2019d, \mn@doi [\apj] {10.3847/1538-4357/ab113b},
  \href {https://ui.adsabs.harvard.edu/abs/2019ApJ...875..122A} {875, 122}

\bibitem[\protect\citeauthoryear{{Abbott} et~al.}{{Abbott}
  et~al.}{2019e}]{2019ApJ...879...10A}
{Abbott} B.~P.,  et~al., 2019e, \mn@doi [\apj] {10.3847/1538-4357/ab20cb},
  \href {https://ui.adsabs.harvard.edu/abs/2019ApJ...879...10A} {879, 10}

\bibitem[\protect\citeauthoryear{{Abbott} et~al.}{{Abbott}
  et~al.}{2020a}]{2020arXiv200714251T}
{Abbott} R.,  et~al., 2020a, preprint, \href
  {https://ui.adsabs.harvard.edu/abs/2020arXiv200714251T} {} (\mn@eprint
  {arXiv} {2007.14251})

\bibitem[\protect\citeauthoryear{{Abbott} et~al.}{{Abbott}
  et~al.}{2020b}]{2020ApJ...892L...3A}
{Abbott} B.~P.,  et~al., 2020b, \mn@doi [\apjl] {10.3847/2041-8213/ab75f5},
  \href {https://ui.adsabs.harvard.edu/abs/2020ApJ...892L...3A} {892, L3}

\bibitem[\protect\citeauthoryear{{Andersson}}{{Andersson}}{1998}]{1998ApJ...502..708A}
{Andersson} N.,  1998, \mn@doi [\apj] {10.1086/305919}, \href
  {http://adsabs.harvard.edu/abs/1998ApJ...502..708A} {502, 708}

\bibitem[\protect\citeauthoryear{{Andersson} \& {Pnigouras}}{{Andersson} \&
  {Pnigouras}}{2020}]{2020PhRvD.101h3001A}
{Andersson} N.,  {Pnigouras} P.,  2020, \mn@doi [\prd]
  {10.1103/PhysRevD.101.083001}, \href
  {https://ui.adsabs.harvard.edu/abs/2020PhRvD.101h3001A} {101, 083001}

\bibitem[\protect\citeauthoryear{{Andersson}, {Kokkotas}  \&
  {Stergioulas}}{{Andersson} et~al.}{1999}]{1999ApJ...516..307A}
{Andersson} N.,  {Kokkotas} K.~D.,   {Stergioulas} N.,  1999, \mn@doi [\apj]
  {10.1086/307082}, \href {http://adsabs.harvard.edu/abs/1999ApJ...516..307A}
  {516, 307}

\bibitem[\protect\citeauthoryear{{Baym} \& {Pines}}{{Baym} \&
  {Pines}}{1971}]{1971AnPhy..66..816B}
{Baym} G.,  {Pines} D.,  1971, \mn@doi [Ann. Phys.]
  {10.1016/0003-4916(71)90084-4}, \href
  {https://ui.adsabs.harvard.edu/abs/1971AnPhy..66..816B} {66, 816}

\bibitem[\protect\citeauthoryear{{Bildsten}}{{Bildsten}}{1998}]{1998ApJ...501L..89B}
{Bildsten} L.,  1998, \mn@doi [\apjl] {10.1086/311440}, \href
  {http://adsabs.harvard.edu/abs/1998ApJ...501L..89B} {501, L89}

\bibitem[\protect\citeauthoryear{{Chugunov} \& {Horowitz}}{{Chugunov} \&
  {Horowitz}}{2010}]{2010MNRAS.407L..54C}
{Chugunov} A.~I.,  {Horowitz} C.~J.,  2010, \mn@doi [\mnras]
  {10.1111/j.1745-3933.2010.00903.x}, \href
  {https://ui.adsabs.harvard.edu/abs/2010MNRAS.407L..54C} {407, L54}

\bibitem[\protect\citeauthoryear{{Cook}, {Shapiro}  \& {Teukolsky}}{{Cook}
  et~al.}{1994}]{1994ApJ...424..823C}
{Cook} G.~B.,  {Shapiro} S.~L.,   {Teukolsky} S.~A.,  1994, \mn@doi [\apj]
  {10.1086/173934}, \href {http://adsabs.harvard.edu/abs/1994ApJ...424..823C}
  {424, 823}

\bibitem[\protect\citeauthoryear{{Douchin} \& {Haensel}}{{Douchin} \&
  {Haensel}}{2001}]{2001A&A...380..151D}
{Douchin} F.,  {Haensel} P.,  2001, \mn@doi [\aap]
  {10.1051/0004-6361:20011402}, \href
  {https://ui.adsabs.harvard.edu/abs/2001A&A...380..151D} {380, 151}

\bibitem[\protect\citeauthoryear{{Friedman} \& {Schutz}}{{Friedman} \&
  {Schutz}}{1978}]{1978ApJ...221..937F}
{Friedman} J.~L.,  {Schutz} B.~F.,  1978, \mn@doi [\apj] {10.1086/156098},
  \href {https://ui.adsabs.harvard.edu/abs/1978ApJ...221..937F} {221, 937}

\bibitem[\protect\citeauthoryear{{Gittins} \& {Andersson}}{{Gittins} \&
  {Andersson}}{2019}]{2019MNRAS.488...99G}
{Gittins} F.,  {Andersson} N.,  2019, \mn@doi [\mnras] {10.1093/mnras/stz1719},
  \href {https://ui.adsabs.harvard.edu/abs/2019MNRAS.488...99G} {488, 99}

\bibitem[\protect\citeauthoryear{{Gittins}, {Andersson}  \&
  {Pereira}}{{Gittins} et~al.}{2020}]{2020PhRvD.101j3025G}
{Gittins} F.,  {Andersson} N.,   {Pereira} J.~P.,  2020, \mn@doi [\prd]
  {10.1103/PhysRevD.101.103025}, \href
  {https://ui.adsabs.harvard.edu/abs/2020PhRvD.101j3025G} {101, 103025}

\bibitem[\protect\citeauthoryear{{Haskell}, {Jones}  \& {Andersson}}{{Haskell}
  et~al.}{2006}]{2006MNRAS.373.1423H}
{Haskell} B.,  {Jones} D.~I.,   {Andersson} N.,  2006, \mn@doi [\mnras]
  {10.1111/j.1365-2966.2006.10998.x}, \href
  {http://adsabs.harvard.edu/abs/2006MNRAS.373.1423H} {373, 1423}

\bibitem[\protect\citeauthoryear{{Hessels}, {Ransom}, {Stairs}, {Freire},
  {Kaspi}  \& {Camilo}}{{Hessels} et~al.}{2006}]{2006Sci...311.1901H}
{Hessels} J.~W.~T.,  {Ransom} S.~M.,  {Stairs} I.~H.,  {Freire} P.~C.~C.,
  {Kaspi} V.~M.,   {Camilo} F.,  2006, \mn@doi [\sci]
  {10.1126/science.1123430}, \href
  {http://adsabs.harvard.edu/abs/2006Sci...311.1901H} {311, 1901}

\bibitem[\protect\citeauthoryear{{Horowitz} \& {Kadau}}{{Horowitz} \&
  {Kadau}}{2009}]{2009PhRvL.102s1102H}
{Horowitz} C.~J.,  {Kadau} K.,  2009, \mn@doi [\prl]
  {10.1103/PhysRevLett.102.191102}, \href
  {https://ui.adsabs.harvard.edu/abs/2009PhRvL.102s1102H} {102, 191102}

\bibitem[\protect\citeauthoryear{{Johnson-McDaniel} \&
  {Owen}}{{Johnson-McDaniel} \& {Owen}}{2013}]{2013PhRvD..88d4004J}
{Johnson-McDaniel} N.~K.,  {Owen} B.~J.,  2013, \mn@doi [\prd]
  {10.1103/PhysRevD.88.044004}, \href
  {http://adsabs.harvard.edu/abs/2013PhRvD..88d4004J} {88, 044004}

\bibitem[\protect\citeauthoryear{{Jones}}{{Jones}}{2002}]{2002CQGra..19.1255J}
{Jones} D.~I.,  2002, \mn@doi [Class. Quantum Gravity]
  {10.1088/0264-9381/19/7/304}, \href
  {https://ui.adsabs.harvard.edu/abs/2002CQGra..19.1255J} {19, 1255}

\bibitem[\protect\citeauthoryear{{Jones}}{{Jones}}{2003}]{2003ApJ...595..342J}
{Jones} P.~B.,  2003, \mn@doi [\apj] {10.1086/377351}, \href
  {https://ui.adsabs.harvard.edu/abs/2003ApJ...595..342J} {595, 342}

\bibitem[\protect\citeauthoryear{{Keer} \& {Jones}}{{Keer} \&
  {Jones}}{2015}]{2015MNRAS.446..865K}
{Keer} L.,  {Jones} D.~I.,  2015, \mn@doi [\mnras] {10.1093/mnras/stu2123},
  \href {https://ui.adsabs.harvard.edu/abs/2015MNRAS.446..865K} {446, 865}

\bibitem[\protect\citeauthoryear{{Lattimer} \& {Prakash}}{{Lattimer} \&
  {Prakash}}{2007}]{2007PhR...442..109L}
{Lattimer} J.~M.,  {Prakash} M.,  2007, \mn@doi [\physrep]
  {10.1016/j.physrep.2007.02.003}, \href
  {http://adsabs.harvard.edu/abs/2007PhR...442..109L} {442, 109}

\bibitem[\protect\citeauthoryear{{Melatos} \& {Payne}}{{Melatos} \&
  {Payne}}{2005}]{2005ApJ...623.1044M}
{Melatos} A.,  {Payne} D.~J.~B.,  2005, \mn@doi [\apj] {10.1086/428600}, \href
  {https://ui.adsabs.harvard.edu/abs/2005ApJ...623.1044M} {623, 1044}

\bibitem[\protect\citeauthoryear{{Osborne} \& {Jones}}{{Osborne} \&
  {Jones}}{2020}]{2020MNRAS.494.2839O}
{Osborne} E.~L.,  {Jones} D.~I.,  2020, \mn@doi [\mnras]
  {10.1093/mnras/staa858}, \href
  {https://ui.adsabs.harvard.edu/abs/2020MNRAS.494.2839O} {494, 2839}

\bibitem[\protect\citeauthoryear{{Ottosen} \& {Ristinmaa}}{{Ottosen} \&
  {Ristinmaa}}{2005}]{75da23d2ccdd4ab1b1b32564cc0df76b}
{Ottosen} N.~S.,  {Ristinmaa} M.,  2005, {The Mechanics of Constitutive
  Modeling}.
Elsevier, United States, \mn@doi{10.1016/B978-0-08-044606-6.X5000-0}

\bibitem[\protect\citeauthoryear{{Owen}}{{Owen}}{2005}]{2005PhRvL..95u1101O}
{Owen} B.~J.,  2005, \mn@doi [\prl] {10.1103/PhysRevLett.95.211101}, \href
  {https://ui.adsabs.harvard.edu/abs/2005PhRvL..95u1101O} {95, 211101}

\bibitem[\protect\citeauthoryear{{Papaloizou} \& {Pringle}}{{Papaloizou} \&
  {Pringle}}{1978}]{1978MNRAS.184..501P}
{Papaloizou} J.,  {Pringle} J.~E.,  1978, \mn@doi [\mnras]
  {10.1093/mnras/184.3.501}, \href
  {http://adsabs.harvard.edu/abs/1978MNRAS.184..501P} {184, 501}

\bibitem[\protect\citeauthoryear{{Payne} \& {Melatos}}{{Payne} \&
  {Melatos}}{2006}]{2006ApJ...641..471P}
{Payne} D.~J.~B.,  {Melatos} A.,  2006, \mn@doi [\apj] {10.1086/498855}, \href
  {https://ui.adsabs.harvard.edu/abs/2006ApJ...641..471P} {641, 471}

\bibitem[\protect\citeauthoryear{{Priymak}, {Melatos}  \& {Payne}}{{Priymak}
  et~al.}{2011}]{2011MNRAS.417.2696P}
{Priymak} M.,  {Melatos} A.,   {Payne} D.~J.~B.,  2011, \mn@doi [\mnras]
  {10.1111/j.1365-2966.2011.19431.x}, \href
  {https://ui.adsabs.harvard.edu/abs/2011MNRAS.417.2696P} {417, 2696}

\bibitem[\protect\citeauthoryear{{Singh}, {Haskell}, {Mukherjee}  \&
  {Bulik}}{{Singh} et~al.}{2020}]{2020MNRAS.493.3866S}
{Singh} N.,  {Haskell} B.,  {Mukherjee} D.,   {Bulik} T.,  2020, \mn@doi
  [\mnras] {10.1093/mnras/staa442}, \href
  {https://ui.adsabs.harvard.edu/abs/2020MNRAS.493.3866S} {493, 3866}

\bibitem[\protect\citeauthoryear{{Smoluchowski} \& {Welch}}{{Smoluchowski} \&
  {Welch}}{1970}]{1970PhRvL..24.1191S}
{Smoluchowski} R.,  {Welch} D.~O.,  1970, \mn@doi [\prl]
  {10.1103/PhysRevLett.24.1191}, \href
  {https://ui.adsabs.harvard.edu/abs/1970PhRvL..24.1191S} {24, 1191}

\bibitem[\protect\citeauthoryear{{Ushomirsky}, {Cutler}  \&
  {Bildsten}}{{Ushomirsky} et~al.}{2000}]{2000MNRAS.319..902U}
{Ushomirsky} G.,  {Cutler} C.,   {Bildsten} L.,  2000, \mn@doi [\mnras]
  {10.1046/j.1365-8711.2000.03938.x}, \href
  {http://adsabs.harvard.edu/abs/2000MNRAS.319..902U} {319, 902}

\bibitem[\protect\citeauthoryear{{Wagoner}}{{Wagoner}}{1984}]{1984ApJ...278..345W}
{Wagoner} R.~V.,  1984, \mn@doi [\apj] {10.1086/161798}, \href
  {http://adsabs.harvard.edu/abs/1984ApJ...278..345W} {278, 345}

\makeatother
\end{thebibliography}


\appendix

\section{Calculating the relaxed shape}
\label{app:Appendix}

In this appendix, we demonstrate that the relaxed configuration that is implied, 
but not calculated, in maximum-mountain calculations 
\citet{2000MNRAS.319..902U} and \citet{2013PhRvD..88d4004J} is calculable. 

Suppose one knows the strain of star B, $\sigma_{i j}(\eta)$ (see 
Fig.~\ref{fig:StarsUCB}; Section~\ref{sec:Mountains}). [This is the case in 
\citet{2000MNRAS.319..902U} and \citet{2013PhRvD..88d4004J}.] From the strain 
tensor it is 
possible to obtain the displacement vector, $\eta^i$, which sources the strain. 

We note the following relations: $\delta \rho$ and $\delta p$ are related 
through the equation of state (\ref{eq:PerturbedEOS}), the perturbed Poisson's 
equation (\ref{eq:PerturbedPoissons}) couples $\delta \rho$ and $\delta \Phi$ 
and (\ref{eq:PerturbedH_i}) links $\delta \rho$, $\delta p$, $\delta \Phi$ and 
$\delta H_i$. Therefore, it follows that if any one of 
$(\delta \rho, \delta p, \delta \Phi, \delta H_i)$ are known, the other 
quantities can, in principle, be calculated. 

We begin with (\ref{eq:PerturbedEulerUCB}). Since we know the strain tensor that 
takes one from star A to star B, we also know $\delta H_i^\text{SB}$. This means 
we have $(\delta \rho_\text{SB}, \delta p_\text{SB}, \delta \Phi_\text{SB})$. It 
is this logic, that enables \citet{2000MNRAS.319..902U} and 
\citet{2013PhRvD..88d4004J} to 
compute the quadrupole moment from just the strain tensor. 

By considering variations between star A (\ref{eq:EulerA}) and star B 
(\ref{eq:EulerB}), we find 
\begin{equation}
    \delta H_i^\text{AB} = - f_i + \nabla^j t_{i j}(\eta). 
\label{eq:PerturbedEulerAB}
\end{equation}
We know $t_{i j}(\eta)$, but not $f_i$ or $\delta H_i^\text{AB}$. However, we 
can obtain $\delta H_i^\text{AB}$. The quantity, $\delta H_i^\text{AB}$, is 
generated by the change in shape from star A to star B. This is described by the 
displacement, $\eta^i$. In particular, the two density fields, $\rho_\text{A}$ 
and $\rho_\text{B}$, are linked through the perturbed continuity equation 
(\ref{eq:PerturbedContinuity}). It, therefore, follows that 
\begin{equation}
    \delta H_i^\text{AB} = \delta H_i^\text{AB}(\eta). 
\end{equation}
We rearrange (\ref{eq:PerturbedEulerAB}) to obtain an expression for the force, 
\begin{equation}
    f_i = - \delta H_i^\text{AB}(\eta) + \nabla^j t_{i j}(\eta). 
\end{equation}
Provided $\eta^i$, we can calculate the force which takes the star from a 
spherical shape (star S) to the relaxed shape (star A). 

Using (\ref{eq:EulerS}) and (\ref{eq:EulerA}), we have 
\begin{equation}
    \delta H_i^\text{SA} = f_i. 
\end{equation}
This determines $\delta H_i^\text{SA}$ and, therefore, also 
$(\delta \rho_\text{SA}, \delta p_\text{SA}, \delta \Phi_\text{SA})$. This means 
one can obtain the shape of the relaxed star, supported by a force, $f_i$, with 
the property that when the force is removed the star obtains a strained 
configuration, according to the displacement vector, $\eta^i$. 


\bsp	
\label{lastpage}
\end{document}